\documentclass[%
 reprint,
 floatfix,
 amsmath,amssymb,
 nofootinbib,
 aps,
 pre,
amssymb,
amsmath,
showpacs,
]{revtex4-2}
\usepackage{graphicx} 
\usepackage{booktabs}
\usepackage{tabularx}
\usepackage{dcolumn}
\usepackage{bm}  
\draft

\usepackage{xcolor} 
\usepackage{physics} 
\usepackage{marginnote} 

\usepackage[english]{babel}
\usepackage[mathlines]{lineno} 
\usepackage{mathtools} 
\usepackage{dsfont}  
\usepackage{outlines} 
\usepackage{etoolbox}
\pretocmd{\eqref}{Eq.~}{}{} 
\usepackage[T1]{fontenc}
\usepackage[bb=boondox,cal=cm,bbscaled=1.0]{mathalfa} 
\usepackage{amsthm}
\newtheorem{remark}{Remark}

\newcommand*{\lse}{LSE (\ref{eqn:ls})}

\renewcommand*{\quote}[1]{\textquotedblleft{#1}\textquotedblright}
\usepackage{hyperref} 
\usepackage[capitalize,nameinlink]{cleveref}
\hypersetup{colorlinks=true,allcolors = blue}
\preprint{APS/123-QED}

\newcommand*{\parabolic}[2]{\(\mathcal{P}\qty({#1},{#2})\)}

\newcommand*{\imag}{\mathrm{i}}
\DeclareMathOperator{\erfccom}{erfc}
\newcommand*{\erfc}[1]{\erfccom\qty({#1})}

\newcommand*{\weber}[2]{D_{{#1}}\qty({#2})}

\newcommand*{\deltaS}[1]{\Delta\,s_{{#1}}}

\DeclareDocumentCommand\innerproduct{ s m g }
{ 
	\IfBooleanTF{#1}
	{ 
		\IfNoValueTF{#3}
		{\vphantom{#2}\left\langle\smash{#2},\smash{#2}\right\rangle}
		{\vphantom{#2#3}\left\langle\smash{#2},\smash{#3}\right\rangle}
	}
	{ 
		\IfNoValueTF{#3}
		{\left\langle{#2},{#2}\right\rangle}
		{\left\langle{#2},{#3}\right\rangle}
	}
}

\newcommand*{\hankel}[1]{H_0^{(1)}\qty({#1})}

\begin{document}
\renewcommand{\sectionautorefname}{Sec.}
\renewcommand{\figureautorefname}{Fig.}

\title{
Solutions of the Lippmann-Schwinger equation for confocal parabolic billiards}
\author{Alberto Ruiz-Biestro}
    \email{a01707550@tec.mx}
\affiliation{Photonics and Mathematical Optics Group, Tecnológico de Monterrey, Monterrey, 64849, México}
\author{Julio C. Gutiérrez-Vega}
    \email{juliocesar@tec.mx}
\affiliation{Photonics and Mathematical Optics Group, Tecnológico de Monterrey, Monterrey, 64849, México}
\date{\today}

\begin{abstract}

We present analytical and numerical solutions of the Lippmann-Schwinger equation for the scattered wavefunctions generated by confocal parabolic billiards and parabolic segments with various $\delta$-type potential-strength functions. The analytical expressions are expressed as summations of products of parabolic cylinder functions $D_m$. We numerically investigate the resonances and tunneling in the confocal parabolic billiards by employing an accurate boundary wall method that provides a complete inside-outside picture. The criterion for discretizing the parabolic sides of the billiard is explained in detail. We discuss the phenomenon of transparency at certain eigenenergies.
\end{abstract}

\maketitle

\noindent A. Ruiz-Biestro and J. C. Guti\'errez-Vega, \textit{Solutions of the Lippmann-Schwinger equation for confocal parabolic billiards}, Phys. Rev. E \textbf{109}, 034203 (2024).
\href{https://doi.org/10.1103/PhysRevE.109.034203}{https://doi.org/10.1103/PhysRevE.109.034203}

\section{Introduction}

The Lippmann-Schwinger equation (LSE) is a second kind Fredholm integral equation fundamental in quantum scattering theory \cite{lippmann1950variational,Sakurai,byron2012mathematics,roman1965advanced}. It connects the incident wave $\phi(\vb{r})$ to the scattered wavefunction $\psi(\vb{r})$ of a system in the presence of an interaction potential $V(\vb{r})$. By incorporating the Green function $G_0(\vb{r},\vb{r'})$, the equation enables researchers to study scattering phenomena analytically and numerically.

An application in which the LSE has proven particularly useful is the scattering of plane waves in two-dimensional (2D) structures, often referred to as \textit{quantum billiards} \cite{TabachnikovBOOK}. In recent years, billiards have become relevant because they represent relatively simple systems that are practical for studying a wide variety of quantum phenomena \cite{GutzwillerBOOK,KozlovBOOK}, including energy spectra, eigenstate distributions, chaos, classical-quantum connections, semiclassical approximations, and electromagnetic/optical/quantum analogies, among many others \cite{vahdani2010intersubband,quantumdot,resonator:fuchss2000scattering,resonator:julio2012tunneling,resonator:ree1999aharonov, resonator:ree2002fractal,waalkens1997elliptic}.

For billiards with arbitrary boundaries $\mathcal{C}$, getting closed analytical solutions of the LSE is practically impossible. But, under special conditions, some symmetric billiards allow solutions of the LSE to be found in terms of the eigenfunctions of the Helmholtz equation in the appropriate coordinate system \cite{pereira2022exact}. For example, Maoli et al. \cite{maioli2018,maioli2019} recently characterized the scattering of circular and elliptical billiards using expansions of Bessel and Mathieu functions, respectively.

When it is impossible to obtain closed analytical results, boundary integral methods represent an efficient alternative to numerically calculate the scattering states of billiards with open or closed boundaries $\mathcal{C}$ and different boundary conditions \cite{berry1984,Roach1,Roach2}. These methods have been developing for many years and have several schemes. In particular, in this paper, we adopt the Boundary Wall Method (BWM) introduced by Da Luz et al. in 1997 \cite{daluz1997quantum,zanetti2008resonant,zanetti2008eigenstates,teston2022}.

The scattering of parabolic surfaces has been studied for quite some time in the scalar and electromagnetic formalism \cite{rice1954,morse1954methods,ivanov1963diffraction,newman1988tm,newman1990plane,hillion1997,graham2011}. To simplify the mathematical treatment of the problem, the parabolic barrier is usually extended infinitely to obtain expressions of the scattered field \cite{morse1954methods,ivanov1963diffraction}.
Similarly, the classical and quantum description of a particle confined in a parabolic billiard has been investigated by several authors \cite{Lopac,Fokicheva,villarreal2021classical}. The eigenstates of the parabolic billiard that satisfy the Dirichlet boundary condition are fully analogous to the longitudinal component of the electric field of the TM eigenmodes in parabolic metallic waveguides \cite{noriega2007mode,Spence,Kenney}. Previous research on parabolic geometries, such as the parabolic quantum dots and parabolic billiards, mainly focused on characterizing the confined eigenstates and energy spectra, with little discussion on scattering. To our knowledge, there is no study of LSE solutions for a confocal parabolic billiard.

In this paper, we characterize the scattering of plane waves by confocal parabolic billiards through the analytical and numerical solution of the LSE for different boundary potentials.  We first solve analytically the LSE for parabolic walls of finite and infinite length employing parabolic cylindrical coordinates $(\xi,\eta)$. For this, we take advantage of the series expansions of the plane wave and the 2D Green function in terms of products of parabolic cylinder functions (PCFs) of integer order, $D_m(\cdot)$, often referred to as Weber functions \cite{morse1954methods,erdelyi,Gradshteyn}. Obtaining closed expressions of the scattered field $\psi(\xi,\eta)$ was possible for some special cases. To corroborate the analytical results numerically, we implemented an accurate BWM \cite{daluz1997quantum}, obtaining an excellent agreement with the analytical predictions.

The discretization of the billiard boundary is not trivial since it is formed by two parabolic segments with different lengths and curvatures intersecting at two right corners. We found that the best way to discretize the boundary is to exploit the symmetry about the $x$-axis, which leads to the problem of fitting the corners. The scattering analysis of the confocal parabolic billiards reveals the presence of resonances that agree with the energies of the confined eigenstates reported by \citet{villarreal2021classical}. Additionally, we verify the concept of transparency and duality discussed by \citet{dietz1995inside}.

\section{Preliminaries}\label{Preliminaries}

We will briefly describe the parabolic geometry and the Lippmann-Schwinger equation to establish notation and provide necessary formulas.

\subsection{Expansions in parabolic cylinder functions}

\begin{figure}[t]
    \centering
    \includegraphics[width=6.5cm]{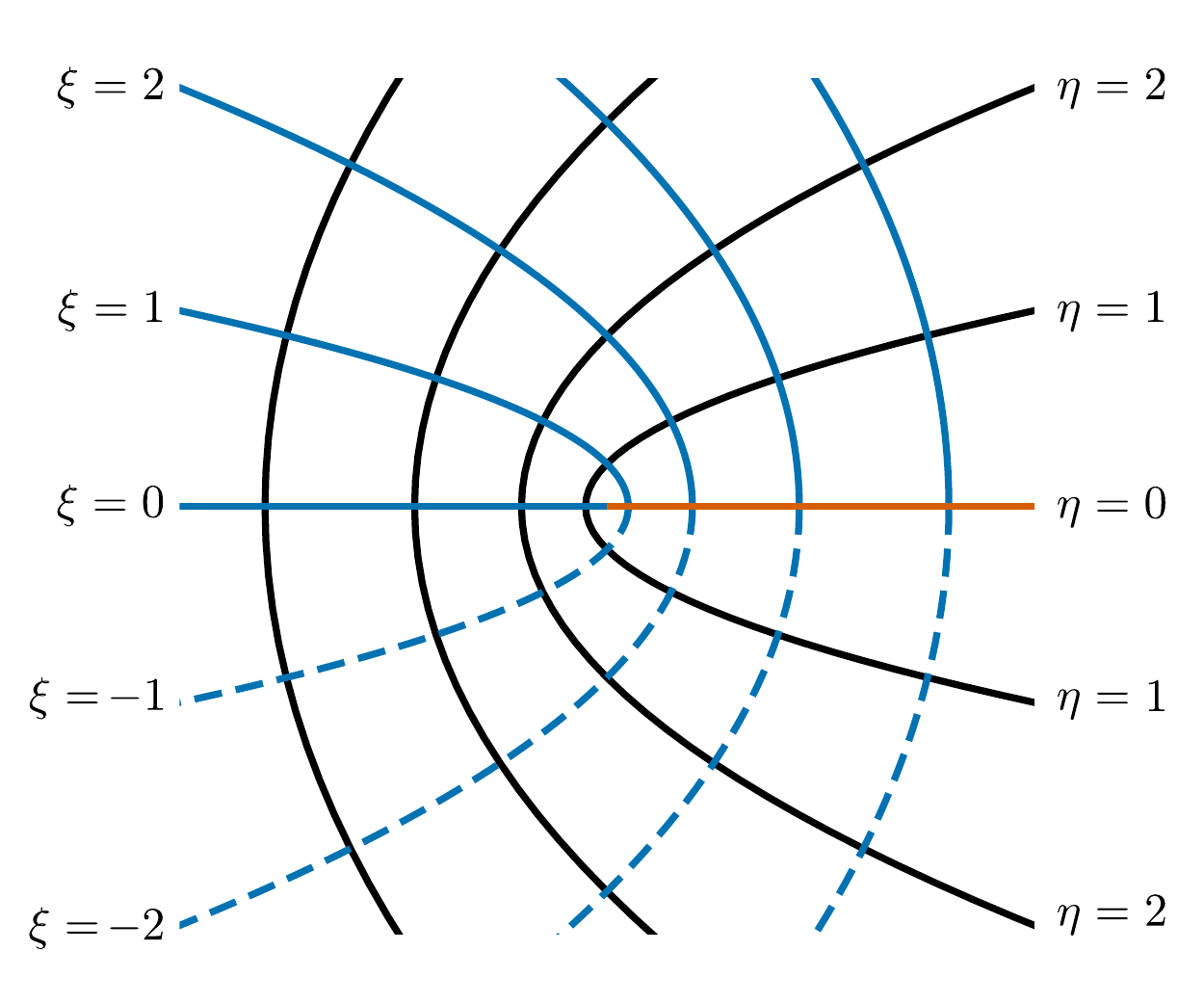}
    \caption{Parabolic cylindrical coordinate system with the branch-cut along the positive $x$-axis. Dashed lines correspond to parabolae with negative $\xi$.}
    \label{fig:parabolas}
\end{figure}

In the transverse plane $\mathbf{r}=(x,y)=(r\cos\theta,r\sin\theta)$, the parabolic cylinder coordinates $\mathbf{r}=(\xi,\eta)$ are related to Cartesian coordinates by
\begin{equation}
    x=\left(\xi^2-\eta^2\right)/2, \qquad y=\xi\eta,
    \label{PCs}
\end{equation}
where $\xi\in(-\infty,\infty)$ and $\eta\in[0,\infty)$. As shown in Fig. \ref{fig:parabolas}, lines of constant $\eta$ and $\xi$ are confocal parabolae opening towards the positive and negative $x$ axis, respectively. In this convention, the $\xi$ coordinate is discontinuous as it crosses the positive $x$ axis; thus, this semi-axis is a branch cut of the coordinate system. This definition is useful for problems where incident plane waves travel mostly from left to right and impinge on parabolic surfaces defined by constant values of $\eta$.
Conversely, if we wanted the negative $x$ axis to be the branch cut, we keep Eqs. (\ref{PCs}) but take the convention $\xi\in[0,\infty)$ and $\eta\in(-\infty,\infty)$.
The scaling factors of the parabolic coordinates are
\begin{equation}
    h_{\xi} = h_{\eta} = h_{\perp} = \sqrt{\xi^2+\eta^2} = \sqrt{2r},
    \label{hh}
\end{equation}
where $r=(x^2+y^2)^{1/2}$ is the radius on the plane $(x,y)$.

The eigenfunctions of the 2D Helmholtz equation in parabolic coordinates, i.e., $(\partial_{\xi}^2+\partial_{\eta}^2) \psi + h_{\perp}^{2} k^2 \psi = 0$, form a complete orthonormal family of wave functions on the transverse plane \cite{morse1954methods,bandres2004parabolic,abramowitz1968}, such that any wave propagating in the plane can be found as an expansion of these eigenfunctions. Due to its relevance in this work, we include the expansions of the plane wave and the zeroth-order Hankel function.

Let
\begin{equation}
    \phi(\mathbf{r};\beta)=\exp[\imag(k_x x + k_y y)] = \exp[\imag k \cos(\theta - \beta)],
    \label{PW}
\end{equation}
be a plane wave with wavenumber $k$ traveling in direction $\beta$,
and $\hankel{k|\vb{r}-\vb{r}'|}$ be the outgoing Hankel cylindrical wave at the observation point $\mathbf{r}$ emitted by a point source located at $\mathbf{r}'$. Their expansions in terms of PCFs $D_m(z)$ are given by \cite{morse1954methods}
\begin{widetext}
\begin{equation}
   \phi(\mathbf{r};\beta) =
   \begin{dcases*}
       \sec\left(\frac{\beta}{2}\right)  \sum_{m=0}^{\infty}%
\frac{\mathrm{i}^{m}\tan^{m}\left(  \beta/2\right)  }{m!}D_{m}\left(\sigma\xi\right)  D_{m}\left(  \mathrm{i}\sigma\eta\right), \qquad & $\left\vert\beta\right\vert <\pi/2$, \\
\csc\left(  \frac{\left\vert \beta\right\vert }{2}\right)
\sum_{m=0}^{\infty}\frac{\mathrm{i}^{m}\cot^{m}\left(  \beta/2\right)}{m!}D_{m}\left(  \mathrm{i}\sigma\xi\right)  D_{m}\left(\sigma\eta\right), \qquad & $\left\vert \beta\right\vert >\pi/2$,
   \end{dcases*}
   \label{PW_Exp}
\end{equation}
and
\begin{equation}
   \hankel{kR} = \frac{\sqrt{8/\pi}}{\imag}
    \begin{dcases*}
        \sum_{m=0}^\infty\dfrac{(-\imag)^m}{m!}[\weber{m}{\sigma\xi'}\weber{m}{\imag\sigma\eta'}]\cdot[\weber{m}{\sigma \xi}\weber{-m-1}{\sigma \eta}],\qquad & $\eta>\eta'$\\
        \sum_{m=0}^\infty\dfrac{(-\imag)^m}{m!}[\weber{m}{\sigma\xi}\weber{m}{\imag\sigma\eta}]\cdot[\weber{m}{\sigma \xi'}\weber{-m-1}{\sigma\eta'}],\qquad & $\eta<\eta'$.
    \end{dcases*}
    \label{eqn:expansion}
\end{equation}
\end{widetext}
where $\sigma\equiv\sqrt{-2\imag k}$, $\imag\sigma\equiv\sqrt{2\imag k}$, and $R=\abs{\vb{r}-\vb{r}'}$.
The primed $(\xi',\eta')$ and non-primed $(\xi,\eta)$ parabolic coordinates specify the location of the source point $\vb{r}'$ and observation point $\vb{r}$, respectively.

The special case of a plane wave traveling into the positive $x$ axis (i.e., $\beta=0$) reduces to
\begin{equation}
    \exp(\imag k x)=\exp\left[ \imag k (\xi^2-\eta^2)/2 \right] = \weber{0}{\sigma \xi}\weber{0}{\imag\sigma\eta},
    \label{PW0}
\end{equation}

In Eqs. (\ref{PW_Exp}) and (\ref{eqn:expansion}), the field $\weber{m}{\sigma\xi}\weber{m}{\imag\sigma\eta}$ is a \textit{well-behaved} wave solution of the 2D Helmholtz equation through the whole plane $(x,y)$ and plays the role of a \emph{standing wave} (like the wavefield $J_m(kr)e^{\imag m \theta}$ in polar coordinates). On the other hand, the field $\weber{m}{\sigma\xi}\weber{-m-1}{\sigma\eta}$ in \eqref{eqn:expansion} plays the role of an outgoing \textit{traveling wave} (like the Hankel wave $H^{(1)}_m(kr)e^{\imag m \theta}$ in polar coordinates). In this way, in \eqref{eqn:expansion}, the series in the exterior region ($\eta>\eta'$) can be interpreted as a summation of outgoing \textit{traveling} waves, while in the interior region ($\eta<\eta'$) as a summation of \textit{standing} waves.

The numerical convergence of the expansions of the Hankel function \eqref{eqn:expansion} is not trivial. Each term of the series involves the product of four PCFs: three having a positive index, $D_m$, and the fourth having a negative index, $D_{-m-1}$. For a given point $(\xi,\eta)$, as $m \rightarrow \infty$, the three functions $D_m$ tend towards zero; but the function $D_{-m-1}$ diverges dramatically. Thus, the three functions $D_m$ and the factorial $m!$ must compensate for the divergent tendency of the function $D_{-m-1}$. The numerical balance is delicate, and high accuracy is needed to evaluate the functions with large $m$. We have verified that the most effective method for evaluating the functions $D_{-m-1}$ is through the use of the recurrence relation \eqref{RCD}, rather than relying on their connection with hypergeometric functions \cite{erdelyi,Gradshteyn}, which is a typical approach taken by commercial software routines for evaluating the PCFs.
In appendix \ref{sec:appendix:weber} , we include a summary of the relevant properties and identities of the PCFs $D_m(z)$ useful for this work \cite{erdelyi,Gradshteyn}.

\subsection{Lippman-Schwinger equation}

The Lippmann-Schwinger equation is a reformulation of the Schr\"odinger equation (satisfying specific boundary conditions) to calculate scattered fields in terms of an integral equation \cite{lippmann1950variational,Sakurai,byron2012mathematics,roman1965advanced}. Consider the scattering of a spinless particle with mass $M$ and free-space wave function $\phi(\vb{r})$ by a potential $V(\vb{r})$ which falls off sufficiently rapidly as $r\rightarrow\infty$. The scattered field $\psi(\vb{r})$ satisfies the Fredholm integral equation ~\cite{roman1965advanced}
\begin{equation}
    \psi(\vb{r})=\phi(\vb{r})+\int\dd[2]{\vb{r}'}\,G_0(\vb{r},\vb{r}')\,V(\vb{r})\,\psi(\vb{r}'),
    \label{eqn:ls}
\end{equation}
where $\phi(\vb{r})$ is an eigenfunction of the unperturbed Hamiltonian, i.e.,  $\hat{\mathcal{H}}_0 (\vb{r})\phi(\vb{r})=E\phi(\vb{r})$, corresponding to the incident wave (typically a plane wave).

In \eqref{eqn:ls}, $G_0(\vb{r},\vb{r}')$ is the two-dimensional Green function for a free-particle with mass $M$, wavenumber $k$, and energy $E=\hbar^2k^2/(2M)$, namely \cite{zangwill2013modern}
\begin{equation} \label{eqn:green}
G_0(\vb{r},\vb{r'};k)=\frac{2M}{\hbar^2}\frac{1}{4i}\hankel{k|\vb{r}-\vb{r}'|},
\end{equation}
where $\hankel{\cdot}$ is the zeroth-order Hankel function of the first kind. This function corresponds to an outgoing circular traveling wave emerging from the point $\vb{r'}$, and it is invariant under the change $\vb{r}\leftrightarrow\vb{r'}$.

We now introduce a boundary-wall potential defined by a continuous sequence of Dirac delta functions along a curve $\mathcal{C}=\mathbf{r}(s)$ parameterized by the variable $s$
\begin{equation}\label{eqn:potential}
V(\vb{r})=\int_{\mathcal{C}}\dd{s}\delta(\vb{r}-\vb{r}(s))\,\gamma(s),
\end{equation}
where $\gamma(s)$ is a potential-strength function. Inserting the $\delta$-potential \eqref{eqn:potential} into the \lse{} gives
\begin{equation}
     \psi(\vb{r}) = \phi(\vb{r})+\int_{\mathcal{C}}\dd{s}'\gamma(s')\,G_0(\vb{r}, \vb{r}(s'))\psi(\vb{r}'),
     \label{eqn:ls:delta}
\end{equation}
where we write $\vb{r}'\equiv\vb{r}(s')$ for simplifying notation. Throughout this paper, we will use primed coordinates for the boundary and unprimed for the observation points.

Replacing the Green function \eqref{eqn:green} into \eqref{eqn:ls:delta} gives
\begin{equation}
\psi(\vb{r}) = \phi(\vb{r})+\alpha\int_{\mathcal{C}}\dd{s}'\gamma(s')\hankel{kR}\psi(\vb{r}'),
\label{eqn:ls:full}
\end{equation}
where $R=|\vb{r}-\vb{r}'|$ and $\alpha\equiv M/(\imag 2\hbar^2)$.
Equation (\ref{eqn:ls:full}) is the starting point to calculate, analytically and numerically, the scattered field of a delta-like boundary $\mathcal{C}$ with the potential-strength function $\gamma(s)$. In the following sections, we will evaluate it for several parabolic boundaries.

\subsection{Boundary Wall Method}\label{sec:bwm}
The BWM was introduced by \citet{daluz1997quantum} as a numerical method for evaluating the solutions of the LSE \eqref{eqn:ls:full}. It was applied in scattering by billiards \cite{zanetti2008eigenstates}, and open resonator structures \cite{resonator:julio2012tunneling, julio2014scalar,resonator:ree1999aharonov,resonator:ree2002fractal}, among others. The method can also be adapted to Neumann and mixed boundary conditions~\cite{daluz1997quantum}. For the sake of brevity, in this subsection, we describe only the fundamental equations of the BWM. In appendix \ref{appBWM}, we include a more detailed description of the BWM for interested readers and reference purposes.

The key principle of the BWM is to calculate $\psi(\vb{r})$ at the boundary $\mathcal{C}$ and apply a single quadrature to obtain the scattered wave in free space.
The boundary is discretized into $N+1$ points $\qty{\mathbf{r}_j \in \mathcal{C}}$, forming $N$ segments with equal lengths preferably.
We then define column vectors for the waves at the boundary, $\Psi=[\psi(\vb{r}_1),\dots,\psi(\vb{r}_N)]^T$ and $\Phi=[\phi(\vb{r}_1),\dots,\phi(\vb{r}_N)]^T$. After this, we solve for the wave function at each boundary segment through
\begin{equation}
    \gamma_i\Psi_i=(\mathbb{T}\Phi)_i=\gamma_i \sum_{j=1}^N\qty[\qty(\mathbb{1}-\mathbb{M}\gamma_j)^{-1}]_{ij}\Phi_{j},
\end{equation}
where $\gamma_i$ is the potential strength at $\mathcal{C}_i$, $\mathbb{1}$ is the identity matrix, $\mathbb{M}$ is a square matrix that contains information about the boundary with entries
\begin{equation}
    \mathbb{M}_{ij}\equiv\int_{\mathcal{C}_j}\dd{s'}G_0(\vb{r}_i,\vb{r}(s')),
\end{equation}
and
\begin{equation}
    \mathbb{T}=\gamma\qty(\mathbb{1}-\mathbb{M}\gamma)^{-1}.
    \label{TTT}
\end{equation}
Finally, the scattered wave for all space can be calculated with
\begin{equation}
    \psi(\vb{r})=\phi(\vb{r})+\sum_{j=1}^NG_0(\vb{r},\vb{r}_j)\deltaS{j}(\mathbb{T}\Phi)_j,
\end{equation}
where we use a mean value approximation for the integral along the curve, and $\deltaS{j}$ is the length of each segment $\mathcal{C}_j$ along the boundary.

\section{Scattering by a parabolic barrier}\label{sec:SPB}

To characterize the scattering by confocal parabolic billiards, we first develop the scattering of plane waves by single parabolic barriers with finite size.

\begin{figure}[t]
    \centering
    \includegraphics[width=0.8\linewidth]{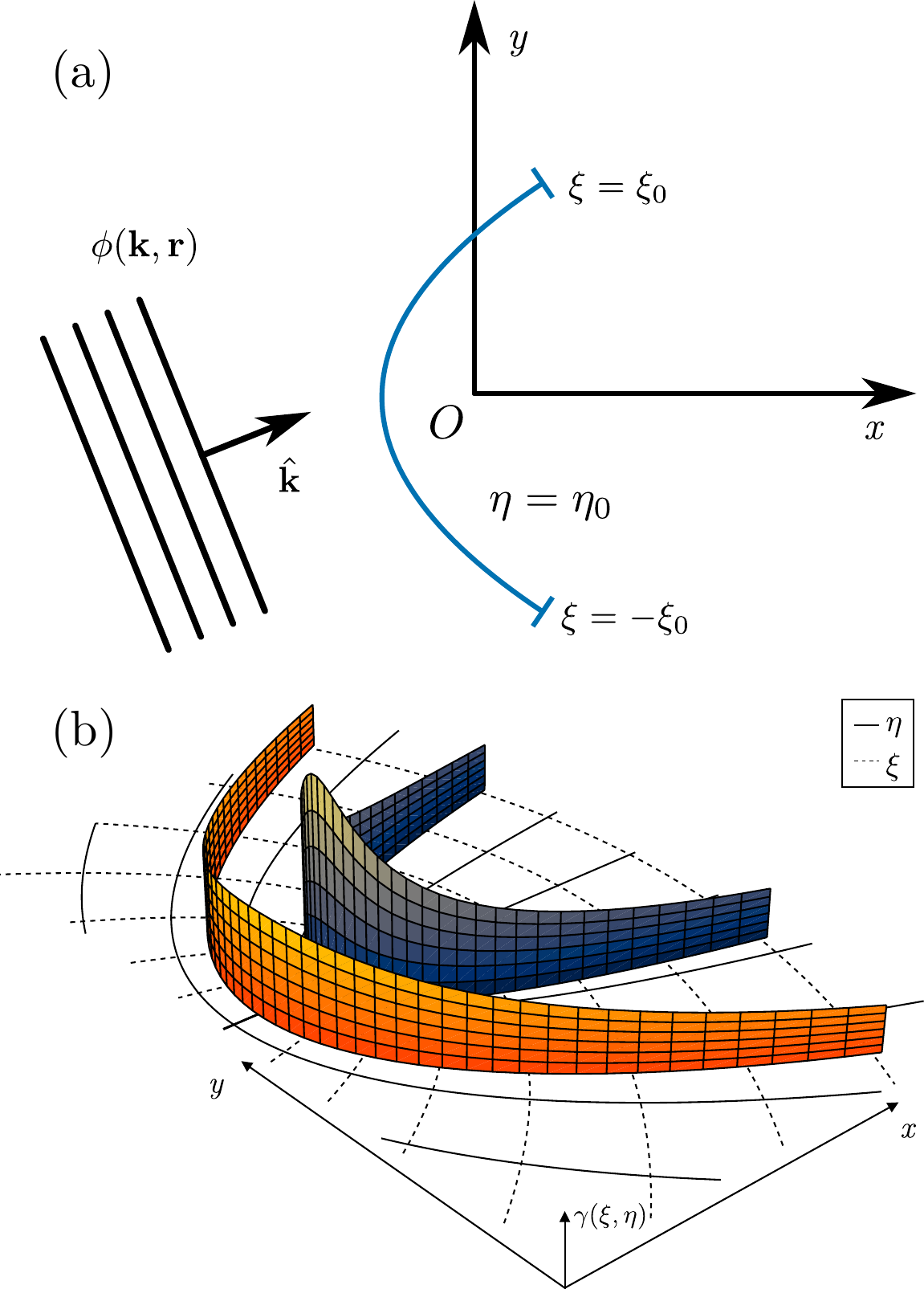}
    \caption{(a) Parabolic barrier with finite size defined for $\xi\in[-\xi_0, \xi_0]$, and $\eta=\eta_0$. A plane wave $\phi(\mathbf{r})=\exp(\imag \mathbf{k}\cdot \mathbf{r})$ in traveling predominantly from left to right. (b) Behavior of the potential $\gamma(s')$ in \eqref{eqn:variable_potential} for two different parabolas, $\eta_0=0.5$ and $\eta_0=1.0$, but the same $\gamma_0 = 1$.}
    \label{fig:parabolic_cylinder}
\end{figure}

Consider a parabolic wall defined by
\begin{equation}
\mathcal{C}=\qty{\vb{r}'=(\xi', \eta_0)~:~\xi'\in(-\xi_0, \xi_0)},
\label{eqn:curve}
\end{equation}
as illustrated in \autoref{fig:parabolic_cylinder}(a). The length of the parabola $\mathcal{C}$ can be easily calculated by integration, we get
\begin{equation}
    L=\xi_0 \sqrt{\xi_0^2 + \eta_0^2} + \eta_0^2 \ln \left[ \frac{\xi_0+\sqrt{\xi_0^2 + \eta_0^2}}{\eta_0} \right].
    \label{length}
\end{equation}

A plane wave $\phi=\exp(\imag \vb{k}\cdot\vb{r})$ travels predominantly from left to right, impinging the convex side of the parabola.
We will refer to the region on the left side of the parabola (its convex side) as the \textit{exterior} region. Conversely, the area on the right side of the parabola (its concave side) will be referred to as the \textit{interior} region.

The LSE (\ref{eqn:ls:full}) for the boundary defined in \eqref{eqn:curve} takes the form
\begin{multline}
    \psi(\vb{r}) = \phi(\vb{r})+\alpha\int\limits_{-\xi_0}^{\xi_0}
    \dd{\xi}'\,\gamma(s')\sqrt{\eta_0^2+\xi'^2}\\\times\hankel{kR}\,\psi(\vb{r}').
    \label{eqn:ls:along_xi}
\end{multline}

For ease of analysis, we rewrite \eqref{eqn:ls:along_xi} as
\begin{equation}
\psi(\vb{r})=\phi(\vb{r})+\alpha\hat{\mathcal{G}}\psi,
    \label{eqn:operator:form}
\end{equation}
where $\hat{\mathcal{G}}$ is an integral operator acting on $\psi$ defined by
\begin{equation}
\label{eqn:operator:definition}
    \hat{\mathcal{G}}\equiv\int\limits_{-\xi_0}^{\xi_0}
    \dd{\xi}'\,\gamma(s')\sqrt{\eta_0^2+\xi'^2}\,\hankel{kR}.
\end{equation}
In the following subsections, we will analyze several realizations of the potential strength function $\gamma(s)$ along the parabolic barrier.

\subsection{Variable potential strength}\label{sec:variable_potential}
We begin by analyzing the case of a potential strength that is proportional to the curvature of the parabolic wall, namely
\begin{equation}
\gamma(s')=\gamma(\xi',\eta_0) = \frac{\gamma_0 \sqrt{\pi/8}}{\sqrt{\eta_0^2+\xi'^2}}
= \frac{\gamma_0}{4} \sqrt{\frac{\pi}{r'}}, \quad\gamma_0\geq 0.
\label{eqn:variable_potential}
\end{equation}
where the factor $\sqrt{\pi/8}$ is introduced for later convenience, and $r'(\xi)$ is the distance  from the origin, see equivalent in \eqref{hh}. The behavior of the delta potential can be appreciated in \autoref{fig:parabolic_cylinder}(b), where we plot it for two different parabolas $\eta_0$. The potential reaches its maximum at the apex of the parabola and decreases monotonically as $|\xi|$ increases. Note in \eqref{eqn:variable_potential} that the potential is inversely proportional to the square root of the radius. The potential \eqref{eqn:variable_potential} simplifies the integral equation \eqref{eqn:operator:definition}, opening the possibility of obtaining closed analytical results. This approach for selecting potential strengths based on the curvature of the scattering wall was also utilized in previous studies, such as \citet{maioli2019}.

By replacing the expansion of the Hankel function $\hankel{kR}$ [\eqref{eqn:expansion}], and inserting the variable potential \eqref{eqn:variable_potential}, the operator $\hat{\mathcal{G}}$ becomes
\begin{multline}
    \hat{\mathcal{G}}= \frac{\gamma_0}{\imag}\sum_{m=0}^\infty \frac{(-\imag)^m}{m!}
    \int\limits_{-\xi_0}^{\xi_0}\dd{\xi'}\,
    [\weber{m}{\sigma\xi'}\weber{m}{\imag\sigma\eta_<}]\\
    \times[\weber{m}{\sigma \xi}\weber{-m-1}{\sigma \eta_>}],
    \label{Eq16}
\end{multline}
where $\eta_{>}$ represents the greater of $\eta$ and $\eta'$, and $\eta_{<}$ the smaller of $\eta$ and $\eta'$.

Since the boundary has constant $\eta_0$, we note the equivalence $\eta'\equiv\eta_0$, then \eqref{Eq16} can be rearranged as
\begin{multline}
\hat{\mathcal{G}} = \frac{\gamma_0}{\imag}\sum_{m=0}^\infty \frac{(-\imag)^m}{m!}
    Q_m(\eta_>,\eta_<)\\
    \times\weber{m}{\sigma\xi}\int\limits_{-\xi_0}^{\xi_0}\dd{\xi'}
        \weber{m}{\sigma\xi'},
\end{multline}
where we define
\begin{equation}\label{eqn:varphi}
    Q_m(u,v)\equiv\weber{-m-1}{\sigma u}\weber{m}{\imag\sigma v}.
\end{equation}

The operator $\hat{\mathcal{G}}$ acting on $\psi$, denoted $\hat{\mathcal{G}}\psi$, suggests using the inner product notation for functions $f$ and $g$
\begin{equation}
\label{eqn:variable:innerproduct}
\ip{f(\xi)}{g(\xi)}\equiv\int\limits_{-\xi_0}^{\xi_0}\dd{\xi}f(\xi)g(\xi),
\end{equation}
to finally express the scattered wave function $\psi(\xi,\eta)$, \eqref{eqn:operator:form}, in terms of PCFs as follows:
\begin{multline}
\label{eqn:sum:parabolic}
  \psi(\xi,\eta) = \phi(\xi, \eta)-
  \imag\alpha\gamma_0
  \sum_{m=0}^\infty
  \Bigg[
  \frac{(-\imag)^m}{m!} c_m(\eta_0) \\
  \times Q_m(\eta_>,\eta_<)\,\weber{m}{\sigma\xi}\Bigg],
\end{multline}
where the coefficients $c_m(\eta_0)$ are given by
\begin{equation}
c_m(\eta_0) \equiv \ip{\weber{m}{\sigma\xi'}}{\psi(\xi',\eta_0)}.
\end{equation}

Equation (\ref{eqn:sum:parabolic}) is the first important result of this paper. It gives the scattered wave $\psi(\xi,\eta)$ produced by a finite parabolic barrier with potential $\gamma(s')$
given by \eqref{eqn:variable_potential}
for an incident wave $\phi(\xi,\eta)$. Obtaining the wave function of the scattered field only requires finding the coefficients $c_m$.

\subsubsection*{Infinitely long parabolic barrier}
We first consider the special case when the parabola extends infinitely, i.e., $\xi_0\rightarrow \infty$. It can be proved using the relation of the PCFs and the Hermite polynomials [\eqref{eqn:weber:hermite}], as well as the orthogonality of Hermite-Gaussian functions that
\begin{equation}\label{eqn:orthogonality}
    \lim_{\abs{\xi_0}\rightarrow\infty}\ip{\weber{m}{\xi_0}}{\weber{n}{\xi_0}}=\sqrt{2\pi}\,n!\,\delta_{mn}.
\end{equation}

By applying the inner product [\eqref{eqn:variable:innerproduct}], $\ip{\weber{n}{\sigma\xi}}{\cdot}$, to \eqref{eqn:sum:parabolic}, and evaluating at the boundary $\eta=\eta_0$, we can solve for $c_m(\eta_0)$ and find the following expression for the expansion coefficients:
\begin{equation}
    c_n(\eta_0)=\frac{\ip{D_n(\sigma\xi')}{\phi(\xi', \eta_0)}}{
    1+\imag\alpha\gamma_0\,\sqrt{2\pi}
        (-\imag)^nQ_n(\eta_0,\eta_0)}.
        \label{eqn:trivial:coefficients}
\end{equation}
Therefore, from \eqref{eqn:sum:parabolic}, the scattered wave by an infinite parabolic barrier with potential strength \eqref{eqn:variable_potential} is given by
\begin{multline}
\psi(\xi,\eta)=\phi(\xi, \eta) - \imag\alpha\gamma_0\\
  \times\sum_{m=0}^\infty
  \Biggl[%
  \frac{(-\imag)^m}{m!}
    Q_m(\eta_>,\eta_<)\weber{m}{\sigma\xi}
    \,\\
    \times\frac{p_m(\eta_0)}{
    1+\imag\alpha\gamma_0\,\sqrt{2\pi}
        (-\imag)^mQ_m(\eta_0,\eta_0)
        }
    \Biggr].
    \label{eqn:solution:coef}
\end{multline}
where $p_m(\eta_0)\equiv\ip{D_m(\sigma\xi)}{\phi(\xi, \eta_0)}$ are expansion coefficients for the incident wave.

Let us consider as incident wave the plane wave propagating in direction $\beta$ given by \eqref{PW}. Applying the orthogonality relationship \eqref{eqn:orthogonality}, the coefficients are
\begin{equation}\label{eqn:p_n}
p_m(\eta_0)=\sqrt{2\pi}\weber{m}{\imag\sigma\eta_0}\sec\qty(\dfrac{\beta}{2})\frac{\imag^m\tan^m\qty(\beta/2)}{m!}.
\end{equation}

If further, we assume that the plane wave travels horizontally coming from the left \eqref{PW0}, we easily check using \eqref{eqn:p_n} that the only nonzero term in the sum \eqref{eqn:solution:coef} occurs for $m=0$, since
\begin{equation}
p_m(\eta_0)=\begin{cases}
    \sqrt{2\pi}\weber{0}{\imag\sigma\eta_0},&m=0\\
    0,&m\neq 0.
    \end{cases}
\end{equation}
Finally, replacing $p_m(\eta_0)$ and \eqref{eqn:varphi} into \eqref{eqn:solution:coef} yields
\begin{multline}\label{eqn:sol:weber}
\psi(\xi,\eta) = \weber{0}{\sigma\xi}\Bigg[\weber{0}{\imag\sigma\eta}\\
        -\frac{%
        \imag\alpha\gamma_0\sqrt{2\pi}\weber{-1}{\sigma\eta_>}\weber{0} {\imag\sigma\eta_<}\weber{0}{\imag\sigma\eta_0}
        }{1+ \imag\alpha\gamma_0\sqrt{2\pi}\weber{-1}{\sigma\eta_0}\weber{0}{\imag\sigma\eta_0}}\Bigg].
\end{multline}
This expression gives the scattered field in both the exterior $(\eta>\eta_0)$ and the interior $(\eta<\eta_0)$ regions of the infinitely long parabolic barrier.

The interior field is particularly interesting because it is proportional to the incident plane wave \eqref{PW0}, namely
\begin{equation}\label{eqn:sol:inside}
\psi(\xi,\eta<\eta_0) = C(\eta_0) \weber{0}{\sigma\xi}\weber{0}{\imag\sigma\eta},
\end{equation}
where
\begin{equation}
    C(\eta_0) = 1 - \frac{\imag\alpha\gamma_0\sqrt{2\pi}\weber{-1}{\sigma\eta_0}\weber{0}{\imag\sigma\eta_0}}{1+\imag\alpha\gamma_0\sqrt{2\pi}\weber{-1}{\sigma\eta_0}\weber{0}{\imag\sigma\eta_0}}.
\end{equation}

\begin{figure}[t]
    \centering\includegraphics[width=0.85\linewidth]{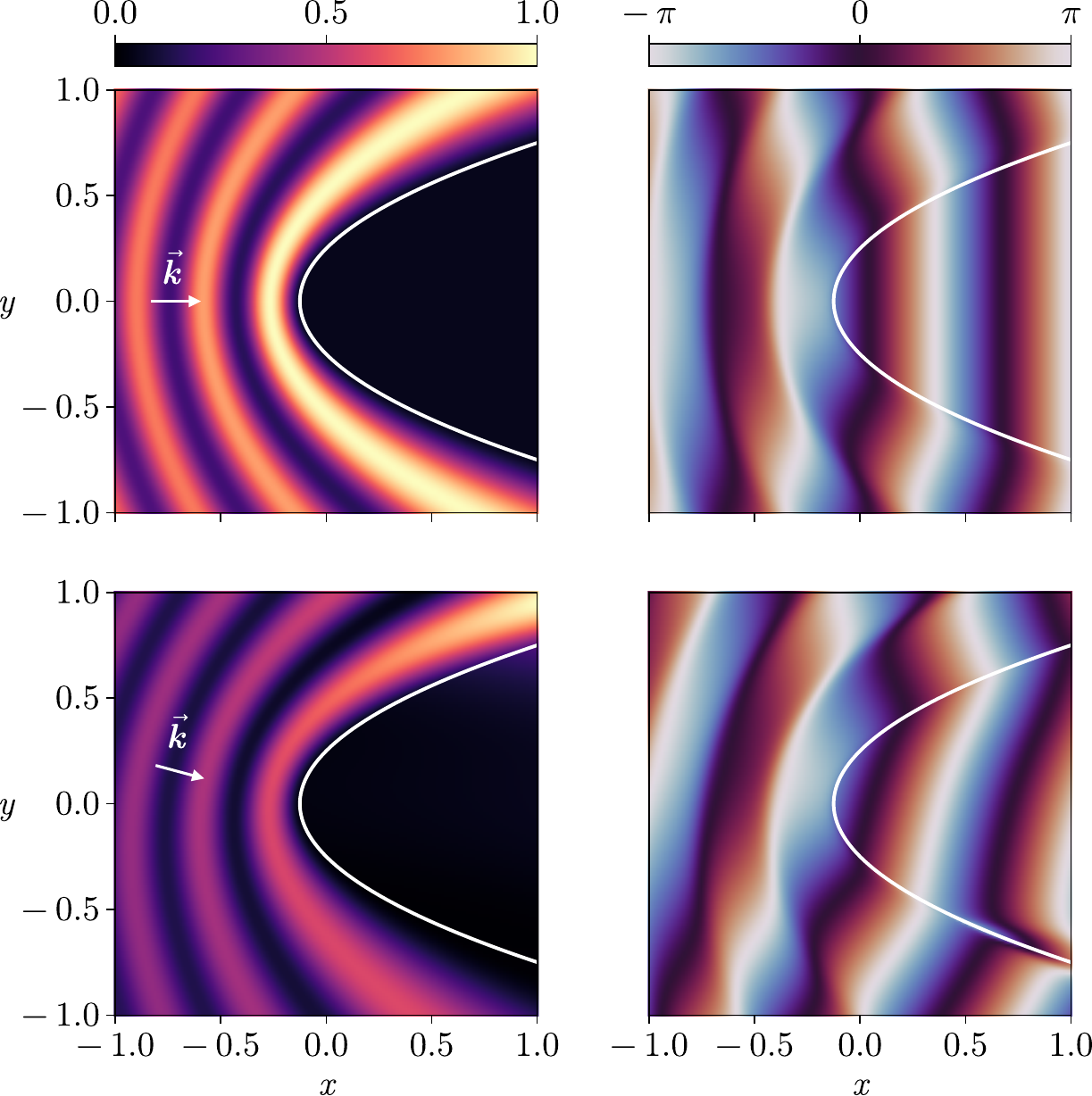}
    \caption{Probability density $\abs{\psi(\xi,\eta)}^2$ and the phase distribution $\arg{\psi(\xi,\eta)}$ of the scattered field by an infinitely-long parabolic barrier ($\eta_0=1/2$) with finite variable strength $\delta$ potential \eqref{eqn:variable_potential} with $\gamma_0=1$. Incident plane wave with $k=10$ at two different incidence angles $\beta=0$ and $\beta = 15^{\circ}$. We use atomic Rydberg units where $\hbar=2M=1$.}
    \label{fig:trivial}
\end{figure}

If we let $\gamma_0\rightarrow \infty$, it is straightforward to see that $\abs{\psi}^2$ satisfies a Dirichlet boundary condition of $\psi(\vb{r})|_{\vb{r}\in\mathcal{C}}=0$, and that the wave-function inside the parabola and at the boundary, $\eta\leq\eta_0$,  vanishes. Furthermore, if we let $\gamma_0\rightarrow 0$, we find $\psi(\xi,\eta)=\phi(\xi,\eta)$, which is what we would expect from no interaction with the boundary, i.e.,
\begin{multline}\label{eqn:limit}
    \lim_{\gamma_0\rightarrow\infty}\psi(\xi,\eta)=\weber{0}{\sigma\xi}\Bigg[\weber{0}{\imag\sigma\eta}\\
    -\frac{\weber{-1}{\sigma\eta_>}\weber{0}{\imag\sigma\eta_<}}{\weber{-1}{\sigma\eta_0}}\Bigg].
\end{multline}

Figure \ref{fig:trivial} shows the probability density $\abs{\psi(\xi,\eta)}^2$ and the phase distribution $\arg{\psi(\xi,\eta)}$ of the scattered wave of an incident plane wave with $k=10$ by a parabolic barrier $\eta_0=1/2$ with delta potential \eqref{eqn:variable_potential}. We include two different angles of incidence to make comparisons. In the case when the incident wave is parallel to the $x$ axis, the field in the concave side of the parabola is small but not zero, and in fact, we can see in its phase the expected plane wavefronts predicted by \eqref{eqn:sol:inside}.

The fields in \autoref{fig:trivial} have been calculated by evaluating the analytical expansions \eqref{eqn:sol:inside}. We additionally implemented the BWM to verify the analytical results and found a strong agreement between them. The greatest difference observed between the analytical $|\psi_a|^2$ and numerical $|\psi_n|^2$ is 0.0085, with an average difference of 0.00067.

\begin{figure}[t]
    \centering
    \includegraphics[width=0.85\linewidth]{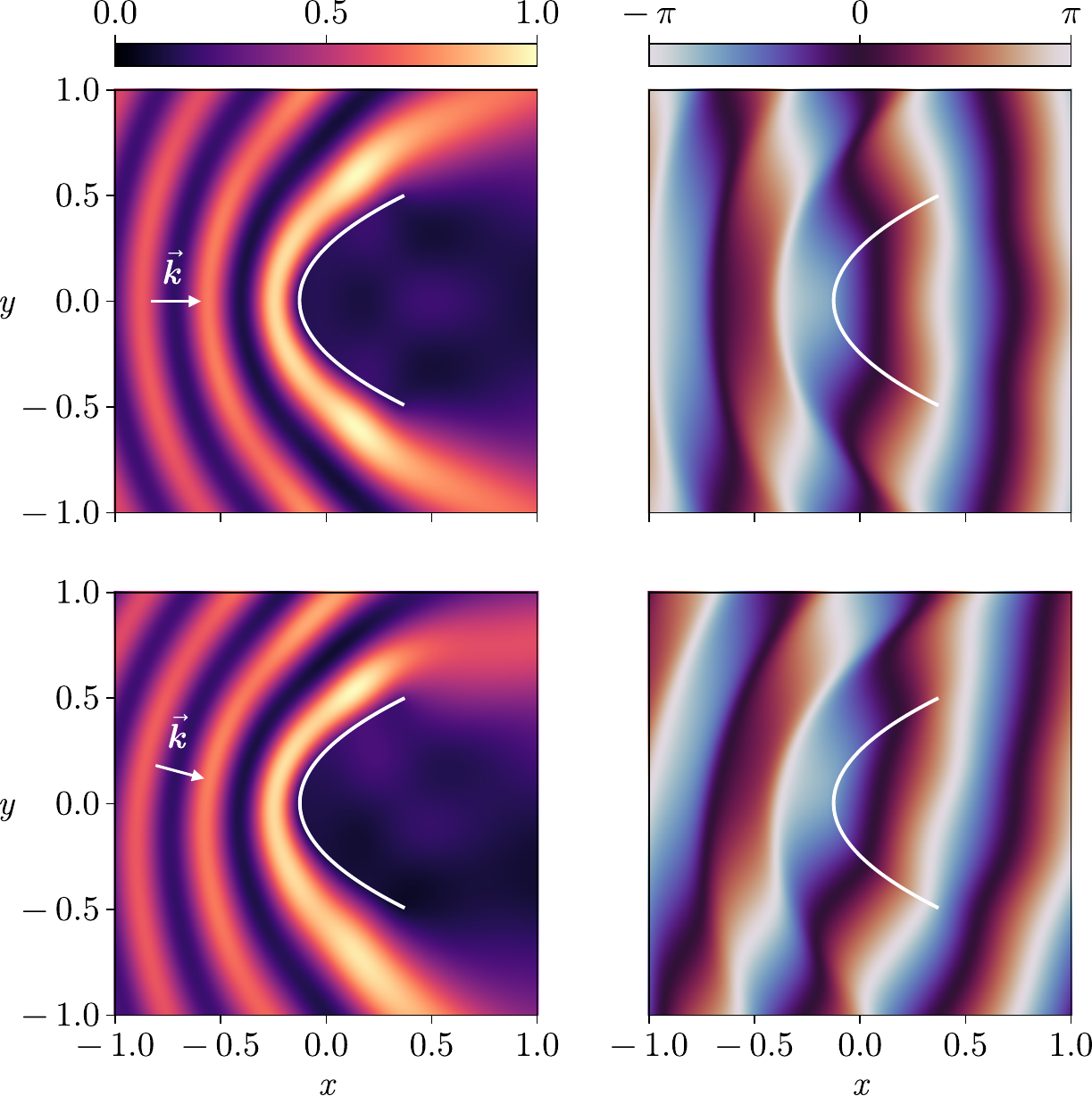}
    \caption{Scattered wave produced by a finite-size parabolic barrier with variable potential-strength $\gamma(s)$ [\eqref{eqn:variable_potential}] with $\gamma_0=1$. The rest of the parameters are those in \autoref{fig:trivial}.}
    \label{fig:infinite_vs_finite}
\end{figure}

\subsubsection*{Finite length barrier.}

Figure \ref{fig:infinite_vs_finite} shows the probability density and phase distribution of the scattered wave produced by a finite-size parabolic barrier with potential-strength function $\gamma(s)$ [\eqref{eqn:variable_potential}] impinged by a plane wave. The parameters of the wall are $\eta_0=1/2$, $\xi\in\qty(-1,1)$, and $\gamma_0 = 1$. The probability density function was calculated by solving numerically \eqref{eqn:sum:parabolic} through the BWM described in detail in the appendix \ref{appBWM} using $N=300$ points to discretize the finite barrier.

\subsubsection*{Knife edge.}

As an additional example, we consider the classical Sommerfeld's diffraction problem of the semi-infinite knife edge, but we consider the potential strength given by \eqref{eqn:variable_potential}. This problem can be seen as the limiting case of the scattering of a plane wave by a parabola when its \textit{parabolicity} goes to zero \cite{sommerfeld1896mathematische,lamb1907sommerfeld,mcdonald2014sommerfeld}. Adopting a similar approach to \citet{lamb1907sommerfeld}, we set $\eta_0=0$ in \eqref{eqn:sol:weber}, consequently, applying the identities of the PCFs in the appendix \ref{sec:appendix:weber}, we have $\weber{-1}{0}=\sqrt{\pi/2}$, $\weber{0}{0}=1$, and $\eta>\eta_0\,\forall\,\eta$. Therefore, for a plane wave with horizontal incidence $\phi = \exp (\imag kx)$, \eqref{eqn:sol:weber} reduces to the simple and closed-form expression
\begin{equation}\label{eqn:knife_edge}
    \psi(\xi,\eta)=\weber{0}{\sigma\xi}\qty[\weber{0}{\imag\sigma\eta}-\frac{\imag\alpha\gamma_0\sqrt{2\pi}\weber{-1}{\sigma\eta}}{1+\imag\alpha\gamma_0\pi}].
\end{equation}

\begin{figure}[t]
    \centering
    \includegraphics[width=0.85\linewidth]{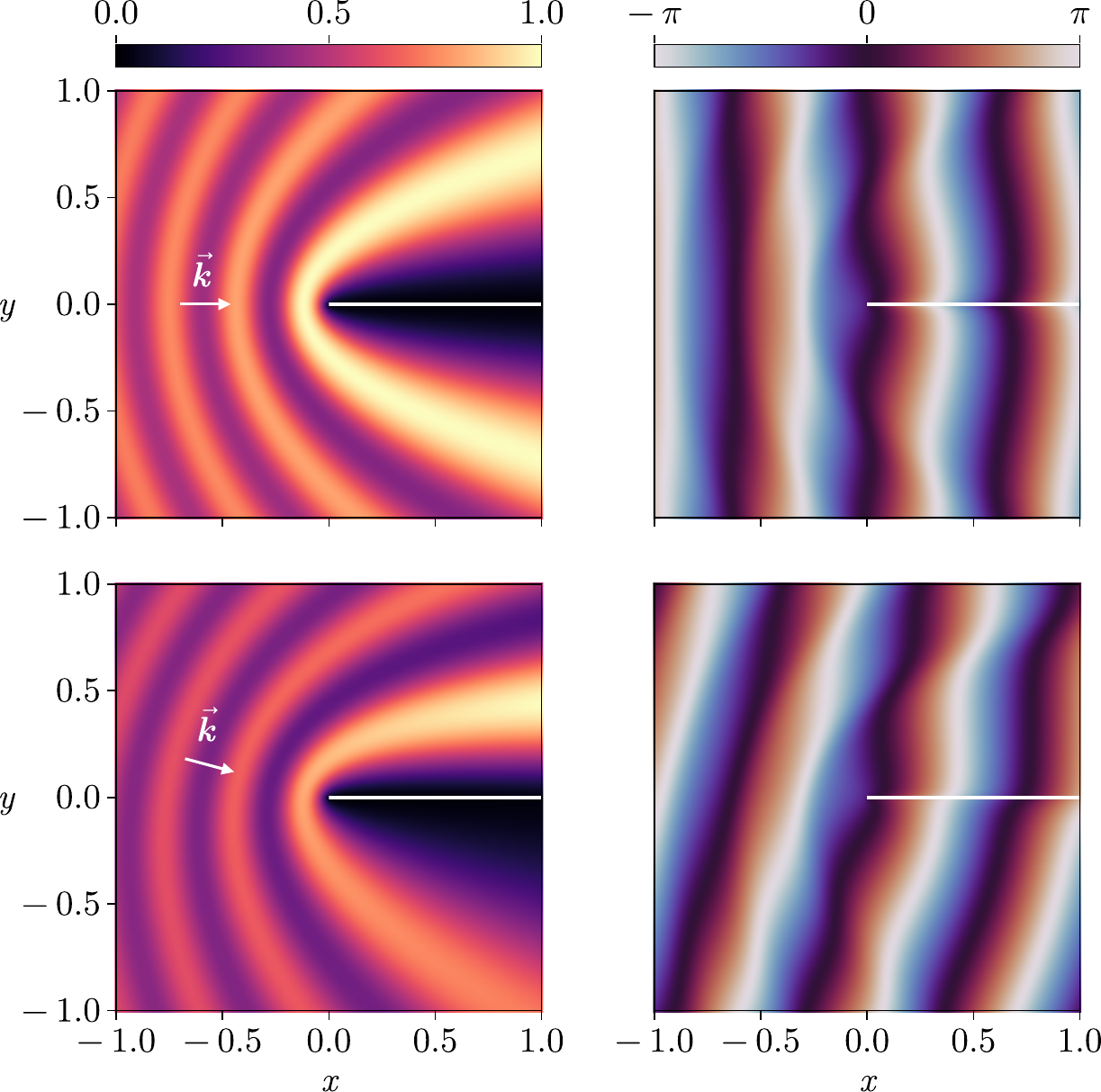}
    \caption{Probability density and phase distribution of the scattered wave by a knife edge with $\gamma(s)$ [\eqref{eqn:variable_potential}] with $\gamma_0=1$. Top row:  Incident plane wave $\exp(\imag 10 x)$. Bottom row:  Incident plane wave $\exp[\imag 10 \cos(\theta+\pi/12)]$.}
    \label{fig:knife_edge}
\end{figure}

Figure \ref{fig:knife_edge}(a) shows the probability density and the phase of the scattered field by a knife-edge for the incident plane wave $\exp(\imag 10 x)$. Both distributions were plotted evaluating directly \eqref{eqn:knife_edge}.  For this symmetric case, the interference fringes between the incident wave and the reflected wave tend to follow the parabolic patterns of the coordinate system with the branch cut on the positive $x$-axis. On the other hand, the phase shows a pattern more similar to that of the incident plane wave.

In contrast to this case, in Fig. \ref{fig:knife_edge}(b), we show the scattered field when the incident plane wave is tilted by 15 degrees. As a result, the fringes of the field lose their symmetry with respect to the knife edge, and the wave reflected from the upper part of the edge becomes more significant than the wave from the lower part. Surprisingly, the phase still maintains a pattern similar to the plane wave but with an inclination of 15 degrees.

Several researchers have conducted studies on this problem with the assumption that the potential of the knife edge is constant \cite{sommerfeld1896mathematische,lamb1907sommerfeld,mcdonald2014sommerfeld}. However, in our case, the potential is variable according to \eqref{eqn:variable_potential}. As a result, the expression (\ref{eqn:knife_edge}) cannot be directly compared with expressions in previous papers.

\subsection{Constant potential strength}\label{sec:constant_potential}

Let us now consider the case of a parabolic barrier [Fig. \ref{fig:parabolic_cylinder}(a)] with constant potential
\begin{equation}
\gamma(s) = \gamma_0 \sqrt{\pi/8}.
\end{equation}
Although a constant potential may seem simpler, the presence of the scale factor $h_{\perp}(\xi,\eta)$ inside the integral operator in \eqref{eqn:operator:definition} makes the method more complicated. The LSE becomes
\begin{multline}
    \psi(\vb{r}) = \phi(\vb{r}) + \alpha\gamma_0 \sqrt{\pi/8}\\
    \times\int\limits_{-\xi_0}^{\xi_0}
    \dd{\xi}'\,\sqrt{\eta_0^2+\xi'^2}\,\hankel{kR}\psi(\vb{r}').
    \label{ppp}
\end{multline}

Again, we replace the expansion of the Hankel function in terms of PCFs \eqref{eqn:expansion} and obtain an expression analogous to \eqref{eqn:sum:parabolic}.
However, there is now a series of inner products [\eqref{eqn:variable:innerproduct}] along the boundary with the functions $f_n(\xi', \eta_0)\equiv\sqrt{\eta_0^2+{\xi'}^2}\,\weber{n}{\sigma\xi'}$.
In other words, we have inner products of the functions $\weber{n}{\sigma\xi'}$ with $f_n(\xi',\eta_0)$.

The inner product in \eqref{eqn:sum:parabolic} had the advantage that, when $\xi_0\rightarrow \infty$, we had an orthogonal relationship that introduced a Kronecker delta, therefore
\begin{multline}
    \psi(\xi,\eta)=
    \phi(\xi, \eta)-\imag\alpha\gamma_0
  \sum_{m=0}^\infty
  \Bigg[\frac{(-\imag)^m}{m!}\\\times
  Q_m(\eta_>,\eta_<)\weber{m}{\sigma\xi}
    \\\times
    \int\limits_{-\xi_0}^{\xi_0}\dd{\xi'}\sqrt{\eta_0^2+\xi'^2}\,\weber{m}{\sigma\xi'}\,\psi(\xi', \eta_0)\Bigg].
\label{eqn:sum:constant}
\end{multline}
Applying the inner product $\ip{f_n(\xi',\eta_0)}{\cdot}$ in \eqref{eqn:sum:constant} and evaluating at the boundary $\eta=\eta_0$, we obtain
\begin{multline}\label{eqn:constant:coefficients}
    c_n(\eta_0)=p_n(\eta_0)-\imag\alpha\gamma_0\sum_{m=0}^\infty\frac{(-\imag)^m}{m!}Q_m(\eta_0, \eta_0)\\\times\,c_m(\eta_0)\,\ip{f_n(\xi', \eta_0)}{\weber{m}{\sigma\xi'}},
\end{multline}
where $c_m(\eta_0)$ and $p_m(\eta_0)$ are the expansion coefficients of the scattered wave $\psi(\xi',\eta_0)$ and the incident wave $\phi(\xi',\eta_0)$, respectively. Note that these coefficients, while equal in notation, differ from those from \eqref{eqn:sum:parabolic}, since our basis is now the set of functions $f_m(\xi',\eta_0)$, instead of PCFs $\qty{\weber{m}{\sigma\xi'}}$. The coefficients are taken at the boundary along $(\xi', \eta')=(\xi',\eta_0)$.

Let us define a matrix operator $\mathbb{F}$ with entries
\begin{equation}\label{eqn:F:def}
\mathbb{F}_{nm}=\frac{(-\imag)^mQ_m(\eta_0,\eta_0)}{m!}\ip{f_n(\xi', \eta_0)}{\weber{m}{\sigma\xi'}},
\end{equation}
where
\begin{equation}
\ip{f_n(\xi', \eta_0)}{\weber{m}{\sigma\xi'}}=\int\limits_{-\xi_0}^{\xi_0}\dd{\xi'}f_n(\xi',\eta_0)\,\weber{m}{\sigma\xi'}.
\end{equation}

We can treat \eqref{eqn:constant:coefficients} as $c_n=p_n-\imag\alpha\gamma_0 \mathbb{F}_{nm} c_m$, which after solving for $c_m$, reads (in matrix form)
\begin{equation}
\label{eqn:system:sol}
\vb{c}=\qty(\mathbb{1}+\imag\alpha\gamma_0\mathbb{F})^{-1}\vb{p}.
\end{equation}
with $\mathbb{1}$ being the identity matrix, and where $\vb{c}$ and $\vb{p}$ are column vectors with elements $c_m(\eta_0)$  and $p_m(\eta_0)$, respectively.

\begin{figure}[t]
    \centering
    \includegraphics[width=0.9\linewidth]{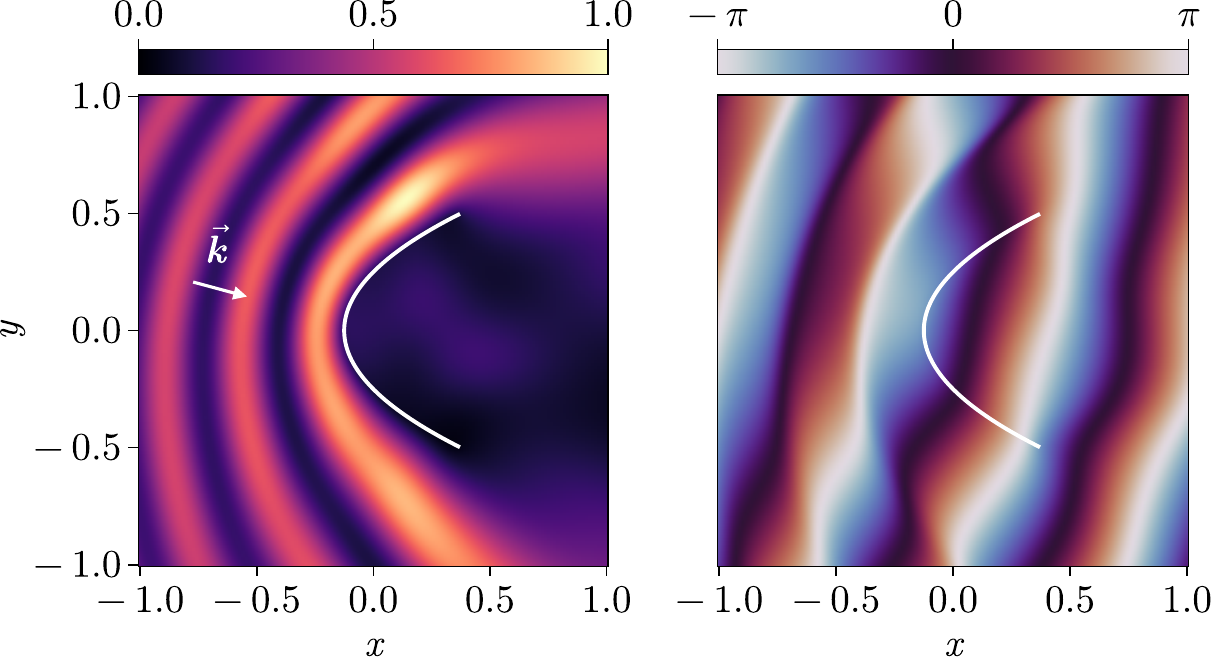}
    \caption{Scattered wave produced by a finite-size parabolic barrier with constant potential strength $\gamma(s)=\gamma_0=10$ and $\eta_0=1/2$, $\xi_0=1$. The rest of the parameters are those in \autoref{fig:trivial}.}
    \label{fig:finite_length_constant}
\end{figure}

The matrix $\mathbb{F}$ is infinite in extent, but for evaluation purposes, we truncate it into a finite $M\times M$ matrix by neglecting elements with smaller contributions. Solving \eqref{eqn:system:sol} should give the coefficients for the expansion \eqref{eqn:sum:constant}. Notice that in \eqref{ppp}, the boundary need not extend to infinity. Therefore, one can solve for the scattering by a finite parabolic segment defined in the interval $(-\xi_0,\xi_0)$. Figure \ref{fig:finite_length_constant} shows the scattered wave generated by a finite-size parabolic barrier with $\eta_0=1/2$, $\xi_0=1$, and constant $\gamma=10$.

\textbf{\textit{Impenetrable} barrier}. A noteworthy simplification occurs when the strength of the potential is infinite, i.e., $\gamma_0\rightarrow \infty$, and the $\delta$-wall becomes \textit{impenetrable}.  From \eqref{eqn:system:sol}, we get
\begin{align}
\lim_{\gamma_0\rightarrow\infty}\gamma_0\,\vb{c}&=\lim_{\gamma_0\rightarrow\infty}\gamma_0\,\qty(\mathbb{1}+\imag\alpha\gamma_0\,\mathbb{F})^{-1}\,\vb{p}, \nonumber\\
    &=(\imag\alpha\mathbb{F})^{-1}\,\vb{p},\nonumber\\
    &=(\imag\alpha)^{-1}\mathbb{W}\vb{p},
\end{align}
where $\mathbb{W} \equiv \mathbb{F}^{-1}$.  We then insert $\gamma_0 \,c_m = (\imag\alpha)^{-1}\,\mathbb{W}_{nm}\,p_m$ into \eqref{eqn:sum:constant} to obtain the scattered wave for an \textit{impenetrable} parabolic barrier with finite size
\begin{multline}
\psi(\xi,\eta)=\phi(\xi,\eta)\\
-\sum_{m=0}^{\infty}\frac{(-\imag)^m}{m!}Q_m(\eta_>,\eta_<)\weber{m}{\sigma\xi} \mathbb{W}_{nm}\, p_m(\eta_0),
\label{eqn:solution:limit}
\end{multline}
This equation is the exact analytical expansion to find the field $\psi$, but we have found that, depending on the parameters, populating and inverting the matrix $\mathbb{F}$ can be numerically more time-consuming and laborious than using the BWM directly. Thus, the BWM provides an easier implementation for finite parabolic barriers in the range $\xi\in(-\xi_0,\xi_0)$, with $\xi_0\ll\infty$.


\section{Scattering by a confocal parabolic billiard}\label{Sec:SCPB}

\begin{figure}[t]
    \centering
    \includegraphics[width=0.7\linewidth]{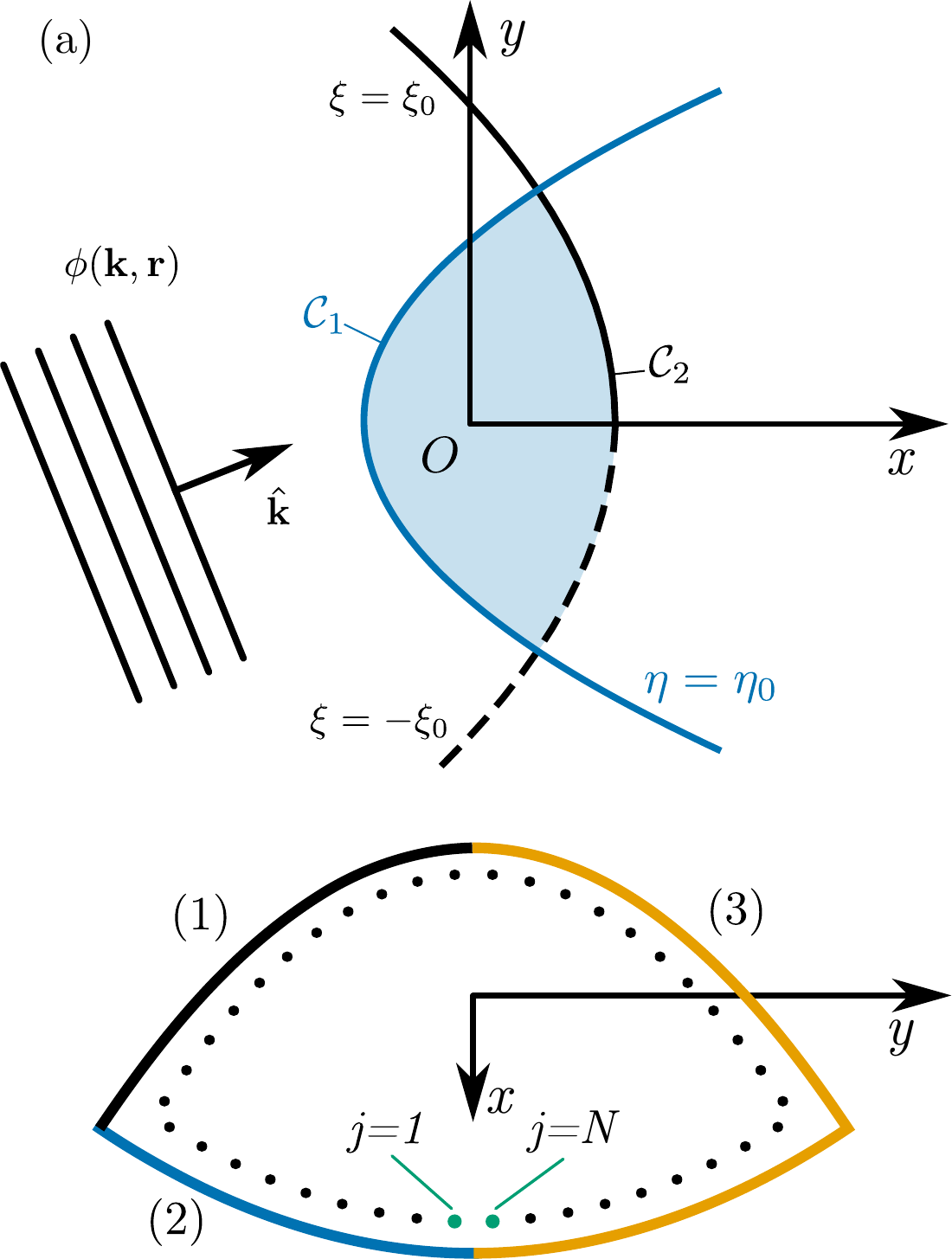}
    \caption{(a) Geometry of the confocal parabolic billiard \parabolic{\xi_0}{\eta_0} formed by the parabolic contours $\mathcal{C}_1$ and $\mathcal{C}_2$. (b) Discretization of the boundary. Dotted line represents midpoints $\vb{r}(s_j)$.}
    \label{fig:concocal_billiard}
\end{figure}

A confocal parabolic billiard consists of two confocal parabolae opening in opposite directions, as shown in \autoref{fig:concocal_billiard}(a). The boundary is defined by the contour $\mathcal{D}=\mathcal{C}_1 \cup \mathcal{C}_2$, with
\begin{subequations}
\begin{align}
\mathcal{C}_1&=\qty{\vb{r}=(\xi, \eta_0):\xi\in[-\xi_0, \xi_0], \xi_0>0},\\
\mathcal{C}_2&=\qty{\vb{r}=(|\xi_0|, \eta):\eta\in[0, \eta_0]}.
\end{align}
\end{subequations}
The potential inside and outside the billiard is equal to zero, and the $\delta$ potential strength at the boundary is given by the functions $\gamma_1(\xi)$ and $\gamma_2(\eta)$ for $\mathcal{C}_1$ and $\mathcal{C}_2$, respectively.

In what follows, we will use the notation \parabolic{\xi_0}{\eta_0} to denote the confocal parabolic billiard, which is defined by intersecting parabolae with parameters $\pm \xi_0$ and $\eta_0$.

From a dynamical point of view, a confocal parabolic billiard is an integrable system whose classical and quantum solutions are known \cite{villarreal2021classical}. A particle confined in the billiard shown in Fig. \ref{fig:concocal_billiard} has two constants of motion: (\textit{i}) the energy and (\textit{ii}) the product of its angular momentum and the momentum in the vertical direction. Action integrals can be found and solved analytically. The eigenstates can also be expressed in terms of parabolic cylinder functions, as can be consulted in Ref. \cite{villarreal2021classical}.

\subsection{Analytic considerations}

Since $\mathcal{C}_1\cap \mathcal{C}_2=\{\varnothing\}$, we can split the integral in the \lse{} into a sum of two integrals over each parabolic boundary. After replacing the potential strength functions $\gamma(s')$ for each segment, we get
\begin{multline}
\psi(\xi,\eta)=\phi(\xi,\eta)+\int\limits_{-\xi_0}^{\xi_0}\dd{\xi}'\,\sqrt{\eta_0^2+\xi'^2}\,\hankel{kR}\,\psi(\xi',\eta_0)\\+\int\limits_{0}^{\eta_0}\dd{\eta'}\,\sqrt{\eta'^2+\xi_0^2}\,\hankel{kR}\,\psi(|\xi_0|,\eta').
\end{multline}

This problem is equivalent to the one of scattering by two potentials, namely $\hat{\mathcal{H}}=\hat{\mathcal{H}}_0+\hat{\mathcal{U}}+\hat{\mathcal{V}}$ \cite[Sec. IV]{two_potential_scattering}, thus
\begin{subequations}
\begin{align}
    \psi_a^{(+)} = \phi_a + \frac{1}{E-H_0+\imag\epsilon}(U+V)\,\psi_a^{(+)},\label{eqn:lse:two} \\
    \psi_a^{(+)}=\chi_a^{(+)}+\frac{1}{E-H_0-U+\imag\epsilon}V\psi_a^{(+)},  \label{eqn:lse:two:chi}
\end{align}
\end{subequations}
where $\phi_a$ is an eigenfunction of $\hat{\mathcal{H}}_0$, i.e., the incident wavefunction.

Equation (\ref{eqn:lse:two}) contains the Green function for both potentials, while \eqref{eqn:lse:two:chi} uses the Green function involving the potential $\hat{\mathcal{U}}$.
The function $\chi$ is related to the incident wave through
\begin{equation}
\chi_a^{(+)}=\phi_a+\frac{1}{E-H_0-U+\imag\epsilon}U\phi_a,
\end{equation}
which shows that one can use its solution in order to solve \eqref{eqn:lse:two:chi}, which is the main problem of scattering by two potentials.

Rewriting the previous equations in the position representation, we get
\begin{subequations}
\begin{equation}
\psi(\vb{r})=\chi(\vb{r})+\int\dd[2]{\vb{r}'}G_U(\vb{r},\vb{r}')\,V(\vb{r}')\,\psi(\vb{r}'),
\end{equation}
where
\begin{equation}
    \chi(\vb{r})=\phi(\vb{r})+\int\dd[2]{\vb{r}'}G_U(\vb{r},\vb{r}')\,U(\vb{r}')\,\phi(\vb{r}),
\end{equation}
\end{subequations}
where $G_U(\vb{r},\vb{r}')$ is the Green function for the potential $U(\vb{r})$, and we have omitted the superscript \quote{$(+)$}  and the subscript \quote{$a$}.

\subsection{Numerical considerations}

To make appropriate comparisons, we will analyze the symmetric \parabolic{2}{2} and the non-symmetric \parabolic{3}{2} parabolic billiards impinged with plane waves traveling at different angles. Actually, the billiard \parabolic{3}{2} will allow us to compare our findings with the results reported by \citet{villarreal2021classical} for the eigenstates of a particle confined in a parabolic billiard with impenetrable walls.

In order to compute the scattered field $\psi(\vb{r})$, we must first discretize the boundary of the billiard.
Placing $N$ points along the edge of a parabolic billiard requires careful consideration. To properly perform discretization, the line segments should all be of equal length. This is not a problem in billiards such as circular or elliptical with smooth borders without corners \cite{maioli2018,maioli2019}. However, parabolic billiards are formed by two contours with different curvatures and lengths, intersecting at corners of 90$^\circ$, as shown in \autoref{fig:concocal_billiard}(a). The calculation of eigenenergies and scattering states is highly sensitive to the point arrangement, so we must use boundary symmetries to arrange them properly.

The discretization scheme we adopted is shown in \autoref{fig:concocal_billiard}(b). First, we discretize half of each parabolic side, e.g., segments (1) and (2), ensuring equally spaced points. The first point of each segment is the closest to the $x$-axis, which is the billiard's symmetry axis. Then, we merge the segments (1) and (2) and mirror them about the $x$-axis to create the symmetric half (3). This procedure ensures that the points are symmetrically placed on the boundary. Since the lengths of $\mathcal{C}_1$ and $\mathcal{C}_2$ are different, the two points surrounding each vertex will not be the same distance apart, nor will they lie right at the corners. This is an unavoidable problem we will have to deal with since we have found that preserving the symmetry and equidistance of the points is more important than forcing the points to fall just at the corners. Of course, a way to reduce the impact of the corner problem is to increase the number of points so that the boundary near the vertices is well sampled. A consequence of our construction method is that the total number of points $N$ is necessarily a multiple of 4.

To set the value of $N$, we need to specify a threshold for the length of each discretization segment. \citet{zanetti2008eigenstates} provided a good \quote{account of the approach accuracy}, which is \quote{given by the number of boundary pieces per wavelength} through
\begin{equation}
    \rho(k)\equiv2\pi N/(\ell k)
    \label{eqn:rho:def},
\end{equation}
where $N$ is the number of segments, $\ell$ the length of the boundary, and $k$ the wavenumber of the incident wave. A value of approximately 10 for $\rho(k)$ is considered reliable, setting a threshold for the number of points, namely
\begin{equation}
N\geq \lceil5\ell k/\pi\rceil.
\label{NN}
\end{equation}

For a billiard \parabolic{3}{2}, the perimeter can be calculated by applying \eqref{length} to both parabolic sides, giving $\ell \approx 28.43$.  For $k$, we consider a maximum value $k=4$. This value will allow us to estimate the first 12 eigenenergies of the billiard \parabolic{3}{2} with infinite walls, according to Ref. \cite{villarreal2021classical}. Replacing $\ell$ and $k$ into \eqref{NN}, we get $N \geq 181$. Because we require $N$ to be a multiple of 4, we will set arbitrarily $N=200$.

\subsection{Resonances and probability distributions}\label{sec:resonances}

Consider a plane wave with wavenumber $k$ incident on a parabolic billiard with high $\delta$-like potential $\gamma$. Depending on its inclination and the dimensions of the billiard, there are certain discrete values of $k$ for which the wave resonates within the billiard. Adopting the usual strategy from previous works \cite{zanetti2008eigenstates,teston2022}, we calculate the resonances by plotting the $L_{1,1}$ norm of the matrix $\mathbb{T}$, i.e., $\norm{\mathbb{T}}\equiv\sum_{i,j}|\mathbb{T}_{ij}|$ as a function of wavenumber $k$. When $k$ gets closer to a resonance $k\rightarrow k_n$, the value of $\norm{\mathbb{T}}$ increases dramatically, forming a delta-like pulse that locates the resonant wavenumber $k_n$.

Computationally speaking, it could be more practical to locate the resonances by monitoring a single element $\mathbb{T}_{i,j}$ rather than evaluating the norm of the entire matrix. However, we have discovered that this method does not always identify all the resonances within the spectral range. Additionally, we have noticed no significant difference in processing time, although the latter approach may perform better regarding memory.

\begin{figure}[t]
    \centering
    \includegraphics[width=\linewidth]{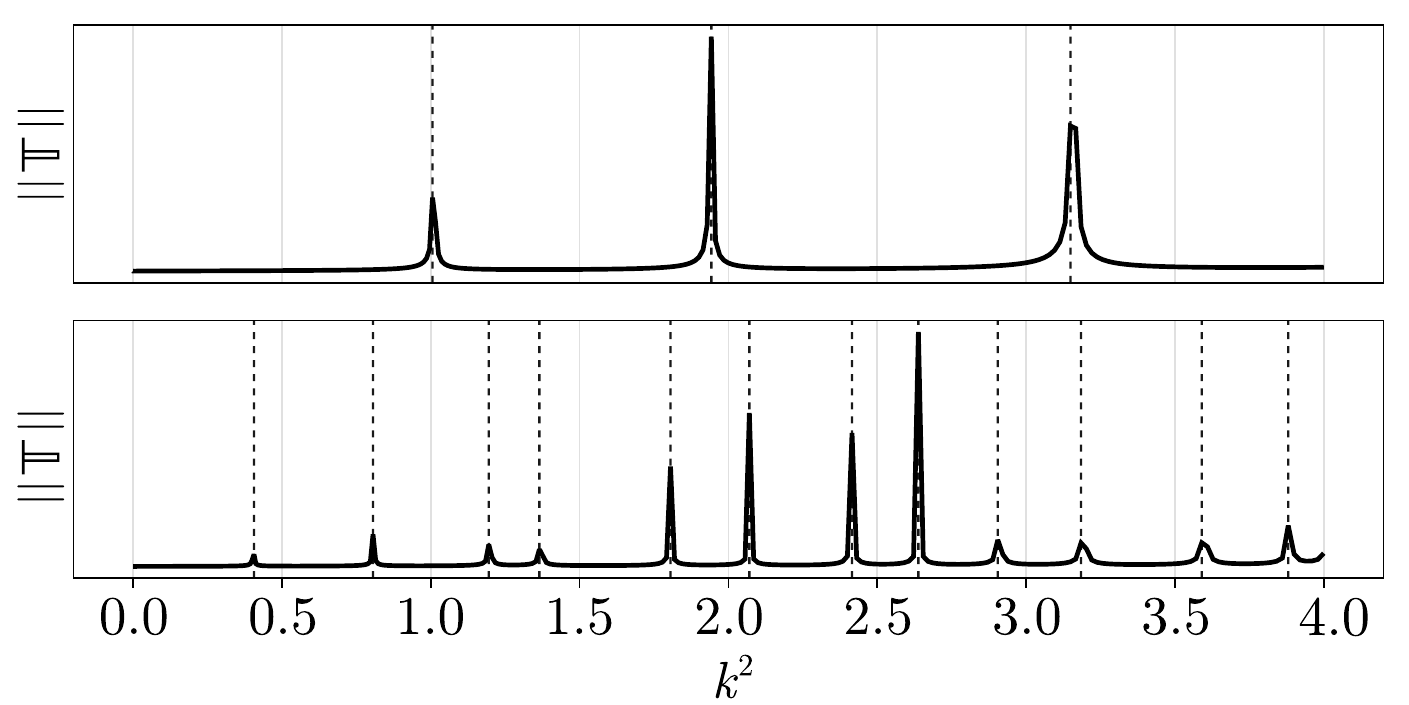}
    \caption{Scan of the energy spectra (in units of $\hbar^2/2M$) of the billiards \parabolic{2}{2} (top) and \parabolic{3}{2} (bottom). Dashed lines indicate the resonances detected by our algorithm. The corresponding field distributions are shown in \autoref{fig:eigenstates}.}
    \label{fig:resonances}
\end{figure}

While the first scan for the \parabolic{3}{2} billiard correctly gives all the resonant states in the range $k^2\in(0,4)$, it may not be the case for any other billiard nor any other range. For example, the \parabolic{2}{2} billiard (\autoref{fig:resonances}, top) shows only one peak in the range $k^2\in(3, 3.5)$. However, upon closer inspection, the iterative algorithm shows that there are actually two peaks, $k^2_n=\qty{3.2876,3.3478}$, which got merged due to the insufficient resolution of the first scan.

\begin{table}[tb]
\caption{Comparison between the first 12 calculated eigenenergies $k_n^2$ for the \parabolic{3}{2} billiard and the reported eigenenergies $k^2_a$ in Ref. \cite{villarreal2021classical}. Both are in units of $\hbar^2/2M$. The fourth column shows the corresponding even (e) or odd (o) stationary eigenstates $\varphi^{m,n}$ shown in \autoref{fig:eigenstates}.}
\footnotesize
\renewcommand{\arraystretch}{1.2}
\centering
\begin{ruledtabular}
\centering
\begin{tabular}{c c c c}
$k_n^2$   & $k_a^2$ & \% error &    $\varphi^{m,n}$\\\hline
0.403776  & 0.403   & 0.078         & $e^{1,1}$ \\
0.807376  & 0.805   & 0.238         & $o^{1,1}$ \\
1.197453  & 1.194   & 0.345         & $e^{2,1}$ \\
1.369032  & 1.365   & 0.403         & $e^{1,2}$ \\
1.802926  & 1.798   & 0.493         & $o^{2,1}$ \\
2.070366  & 2.064   & 0.637         & $o^{1,2}$ \\
2.412215  & 2.405   & 0.721         & $e^{3,1}$ \\
2.637553  & 2.629   & 0.855         & $e^{2,2}$ \\
2.908871  & 2.900   & 0.887         & $e^{1,3}$ \\
3.191544  & 3.181   & 1.054         & $o^{3,1}$ \\
3.596454  & 3.586   & 1.045         & $o^{2,2}$ \\
3.883222  & 3.872   & 1.122         & $o^{1,3}$ \\
\end{tabular}
\end{ruledtabular}
\label{tab:resonances}
\end{table}

Figure \ref{fig:resonances} shows $\norm{\mathbb{T}}$ in function of $k^2$ for the billiards \parabolic{2}{2} (top) and \parabolic{3}{2} (bottom). First, we scan the spectral range with 300 values of $k$, showing us the curve's general behavior. Once we locate the approximate position of a resonance $k_n$, we refine its value by scanning with higher resolution within the spectral window $[k^t_n-\Delta k, k^t_n+\Delta k]$ until satisfying the convergence criterion $\abs{k^t_n-k_n^{t+1}}\leq \epsilon$, where $t$ denotes the $t$-th iteration.  The iterative process is not strictly necessary, but to resolve all the peaks, we need to have high resolution and, therefore, a high number of points, many of which are wasted in plateaus without any resonant states. We remark that although the iterative process is robust, we may need to adjust continuously the peak-finding algorithm to refine the spectrum.


The values of the first 12 resonances depicted in \autoref{fig:resonances} are tabulated in \autoref{tab:resonances}. These correspond to the first 12 scattered fields $\psi_n(\vb{r})$ for the \parabolic{3}{2} billiard. To have a point of comparison, we also tabulated the eigenenergies reported by \citet[Table I]{villarreal2021classical} for the case of the billiard \parabolic{3}{2} with infinite walls, and that were computed by finding the zeroes of the analytic eigenstates in the billiard.

\begin{figure}[t]
    \centering
    \includegraphics[width=\linewidth]{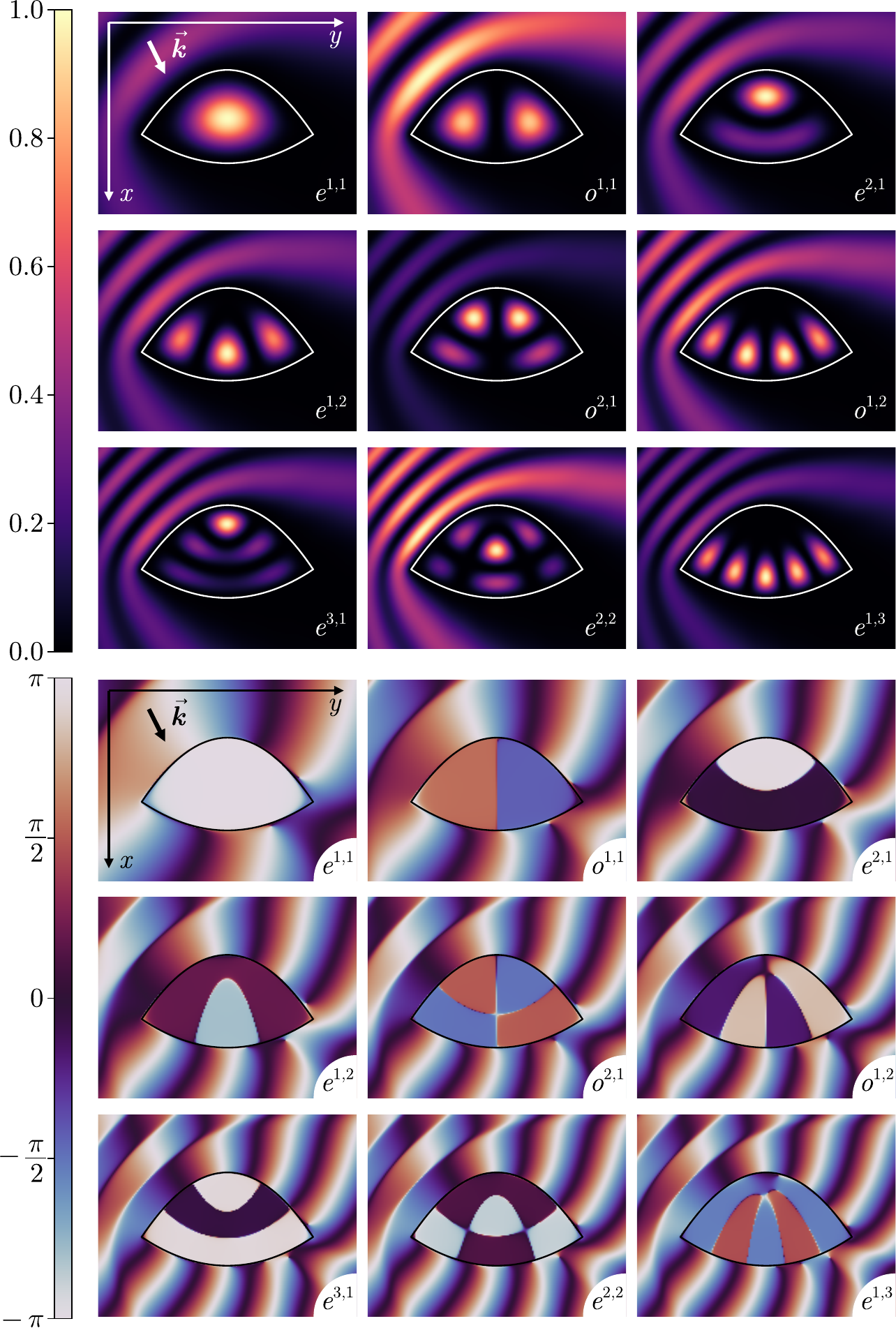}
    \caption{Probability densities $\abs{\psi_n(\vb{r})}^2$ and the phase distributions $\arg{\psi_n(\vb{r})}$ of the first nine scattered wavefields by the \parabolic{3}{2} billiard impinged by a plane wave. The amplitude of all densities has been normalized to unity to facilitate the visualization.}
    \label{fig:eigenstates}
\end{figure}

Figure \ref{fig:eigenstates} shows the probability densities $\abs{\psi_n(\vb{r})}^2$ and the phase distributions $\arg{\psi_n(\vb{r})}$ of the scattered fields corresponding to the eigenenergies $k_n$ in \autoref{fig:resonances} and \autoref{tab:resonances} for a \parabolic{3}{2} billiard. The incident plane wave is traveling in the direction of the unit vector {\small $\widehat{\mathbf{k}}=[2,1]/\sqrt{5}$}. Note that the phase inside the billiard is almost binary, which means that the field inside the cavity is practically a real function, i.e., the fields resemble the interior eigenstates of a point particle trapped in the parabolic billiard \cite{villarreal2021classical}.

The eigenenergies shown in \autoref{fig:resonances} were obtained by discretizing the billiard's boundary using 200 points. However, to achieve greater precision when calculating each distribution $\psi_n(\vb{r})$ shown in \autoref{fig:eigenstates}, we increased the number of points to $N=1200$. This increase in points did not result in a significant increase in computation time, as we only needed to evaluate it once for the particular wavenumber $k_n$. Additionally, a $1200 \times 1200$ matrix can still be inverted with high precision using standard routines available in commercial software.

\begin{figure}[t]
    \centering
    \includegraphics[width=0.9\linewidth]{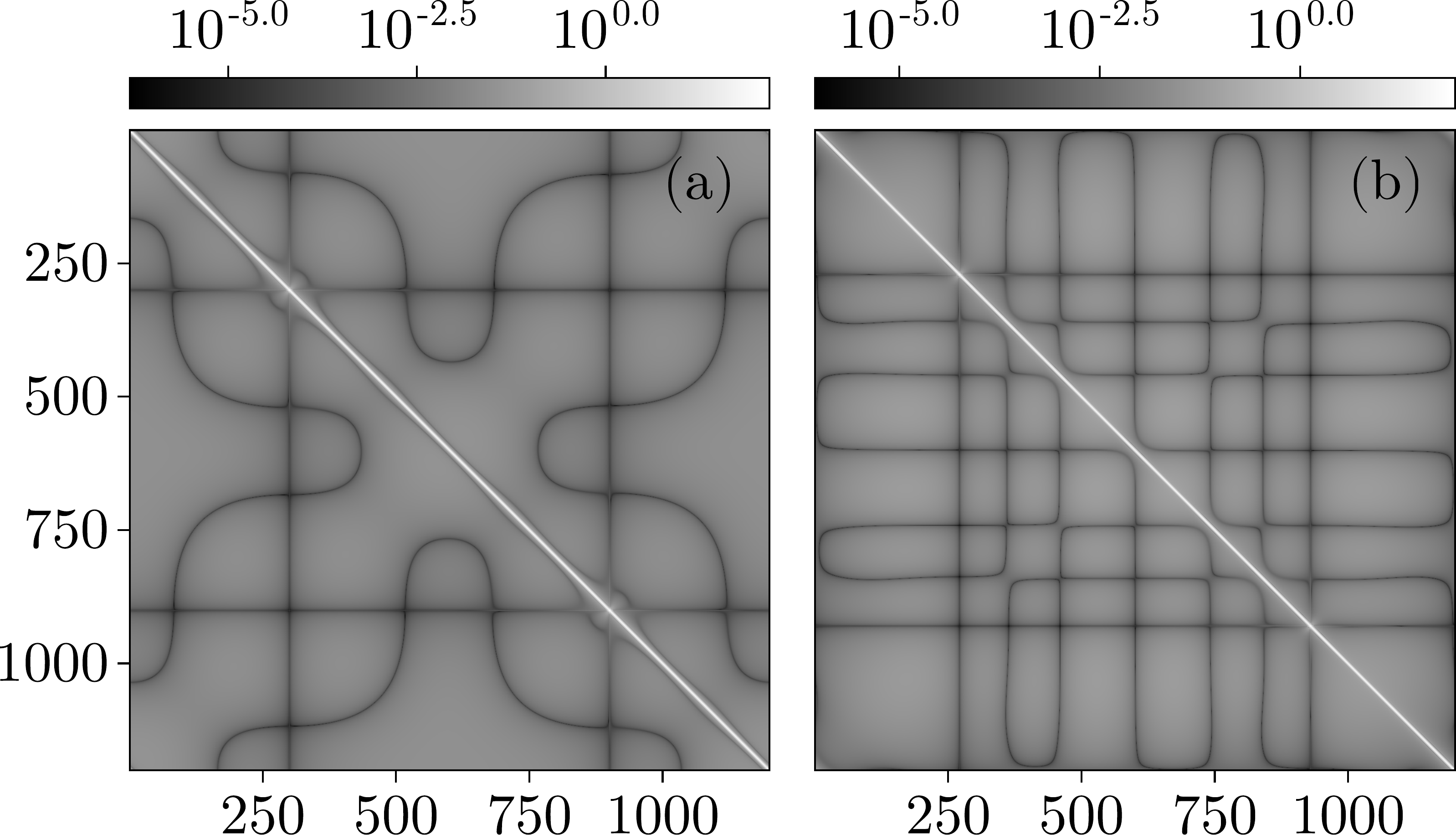}
    \caption{Discretized ($1200\times1200$) density plots of $\abs{\mathbb{T}}$ with $(k_{10})^2=3.181$ for (a) \parabolic{2}{2}, and (b) \parabolic{3}{2} billiards.}
    \label{fig:tmatrix}
\end{figure}

The structure of $\abs{\mathbb{T}}$ is shown in \autoref{fig:tmatrix} for the billiards \parabolic{2}{2} and \parabolic{3}{2} with $k^2=k^2_{10}=3.181$ corresponding to the tenth energy in \autoref{tab:resonances}.  The images display a peculiar arrangement of symmetrical dark lines about the central axes of the matrices. These patterns vary for different values of $k_n$, but remain constant under changes of $N$. If we increase the value of $N$, the dark lines become more defined. Higher frequencies of the scattered fields require finer-scale structure for $\mathbb{T}$ to be characterized accurately, especially since the matrix representation for $G_0(s,s')$ is dense. For the case of the distributions in \autoref{fig:eigenstates}, we evaluated the matrices $\mathbb{T}$ with \eqref{eqn:t:limit} corresponding to an infinite delta-like simplification for the boundary.

In \autoref{fig:tmatrix}, we observe that the matrix of the symmetrical billiard exhibits dark curved lines, which are not present in the matrix of the non-symmetrical one. Further details regarding this dissimilarity in the behavior of the matrices are discussed in Appendix \ref{AppC}.
For a detailed discussion of the $\mathbb{T}$-matrix properties, we refer to Ref. \cite{edwards1998wavelet}.

\begin{figure}[t]
    \centering
    \includegraphics[width=7.5cm]{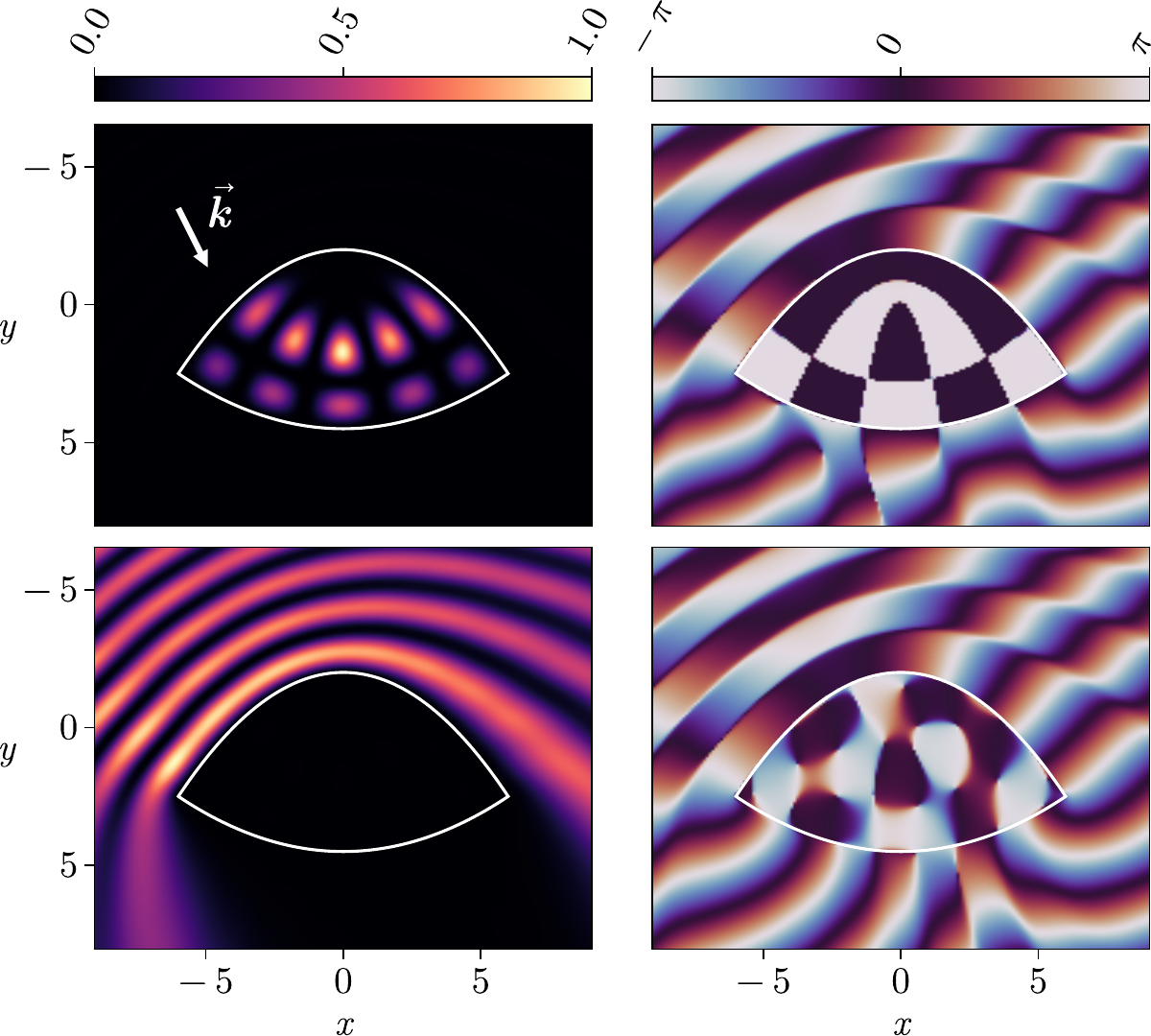}
    \caption{Scattering by the billiard \parabolic{3}{2} with high potential $\gamma_0$ of (top) a resonant plane wave with $k=k_{15}=2.17$, and (bottom) a non-resnonant one with $k=2.19$.}
    \label{fig:comparison_3_2}
\end{figure}

\vspace{3mm}
\textbf{Non-resonant states.}
A slight change in the wavenumber $k$ of the incoming wave $\phi$ with respect to the eigenfrequencies $k_n$ of the billiard can prevent the wave from resonating inside the cavity. If the boundary potential is high, the external wave will not be able to penetrate the billiard. We illustrate this result in \autoref{fig:comparison_3_2} where we plot the scattered field $\psi(\mathbb{r})$ for the resonant wavenumber $k=2.17$ corresponding to the interior eigenstate $_e\psi_{1,5}$ and a nearby wavenumber $k=2.19$.

In the non-resonant case, we can observe that the plane wave surrounds the billiard, generating edge waves and a shadow behind it. Within the billiard, the phase distribution of the resonant wave is binary (i.e., the field is real and stationary), while for the non-resonant wave, it is smooth (i.e., the field is complex and traveling).

It turns out that the \emph{resonances} can be considered as a side-product of a \emph{transparency} phenomenon at the eigenenergy $E_n = k^2_n$. This concept was originally discussed theoretically by \citet{doron1992} and numerically by \citet{eckmann1995} and Dietz et al.~\cite{dietz1995inside} in the context of planar billiards. Basically, at the eigenenergies, the obstacle is transparent in the eyes of the incident wave $\phi$, which translates to the wave function $\psi$ inside the billiard to be the eigenfunction $\psi_n$. Meanwhile, outside the billiard, the wave function corresponds to a scattering matrix $S$ with eigenvalue equal to 1~\cite{dietz1995inside,zanetti2008eigenstates}. \citet{zanetti2008eigenstates} corroborated that the billiard is \quote{transparent} to $\phi$ whenever an eigenvalue of $S(k)$ equals unity, and conversely, if the eigenvalues of $S(k)$ not exactly equal to 1, all $\phi$ with $k\neq k_n$ will be scattered by the billiard.

\begin{figure}[t]
    \centering
    \includegraphics[width=7.5cm]{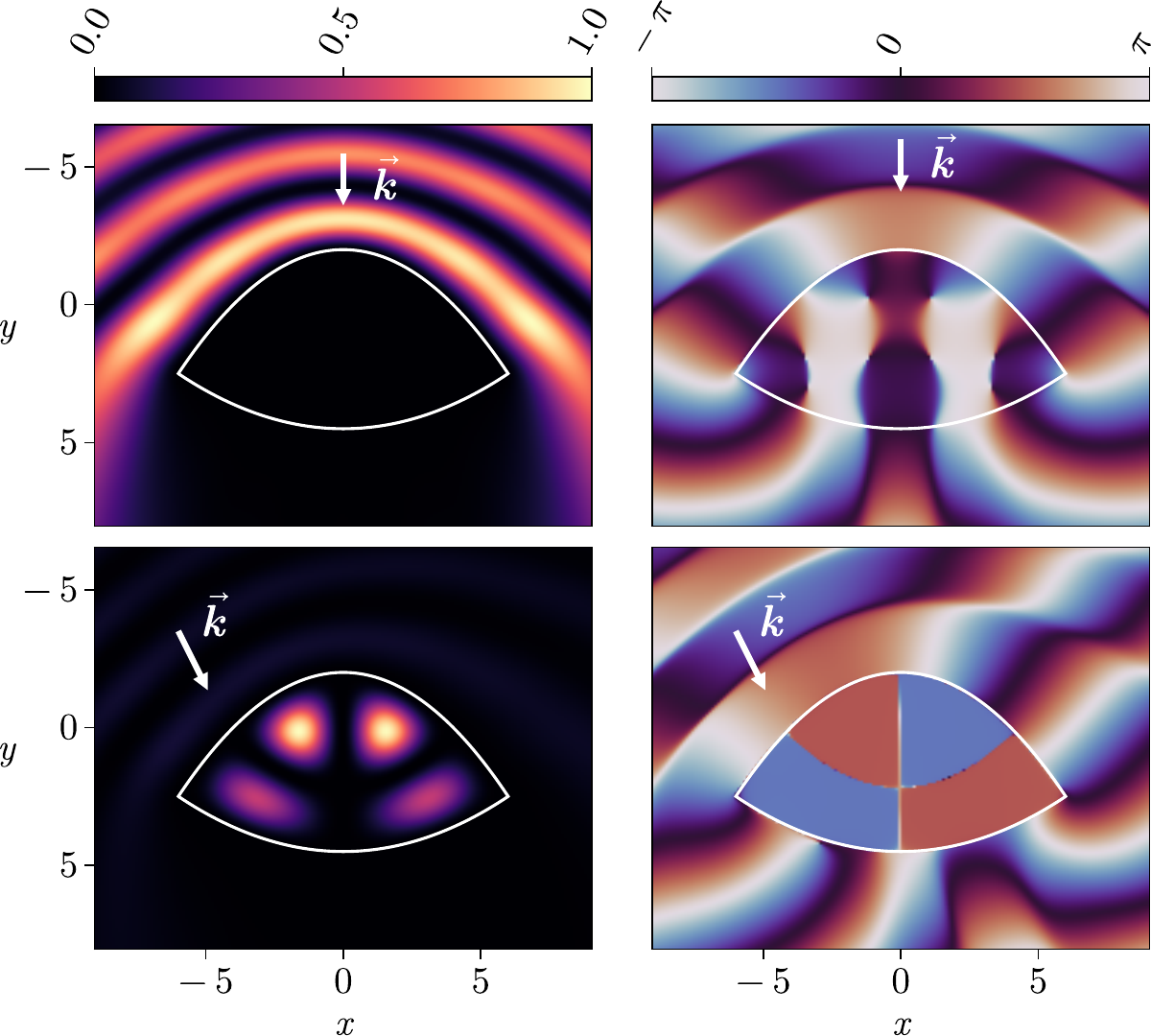}
    \caption{Scattering of the resonant plane wave with $k^2=1.798$ corresponding to the eigenstate $o^{2,1}$ of the billiard \parabolic{3}{2} using two different incident angles, $\vb{k}_{x,y}=[1,0]^T$ (top) and $\vb{k}_{x,y}=[2,1]^T$ (bottom). }
    \label{fig:transparent}
\end{figure}

\vspace{3mm}
\textbf{Symmetrical incidence with respect to the billiard.}
As previously noted in Refs. \cite{dietz1995inside,zanetti2008eigenstates}, resonance in 2D billiards with certain symmetries also depends on the angle of incidence of the plane wave, even when the wavenumber $k$ matches an eigenfrequency $k_n$.
Figure \ref{fig:transparent} illustrates this situation. Let us set $k^2=1.798$ corresponding to the eigenenergy of the odd eigenstate $_o\psi_{2,1}$ of the billiard \parabolic{3}{2} with infinite-strength $\delta$-walls (see Table I and Ref. \cite{villarreal2021classical}). The interior eigenfunction $_o\psi_{2,1}$ is anti-symmetrical about the $x$-axis. As shown in \autoref{fig:transparent} (top), when the incidence is parallel to the $x$-axis, the wave does not resonate within the billiard.
The reason is that the plane wave is symmetrical with respect to the billiard and is incident just along the billiard's axis of symmetry. In order to induce the anti-symmetrical mode within the billiard, we need to break the symmetry of the problem. To achieve this, we cause a tilt in the plane wave so that the incident field at the top $(x>0)$ and bottom $(x<0)$ half-planes are out of phase, as depicted in \autoref{fig:transparent} (bottom).

\begin{figure}[t]
    \centering
    \includegraphics[width=\linewidth]{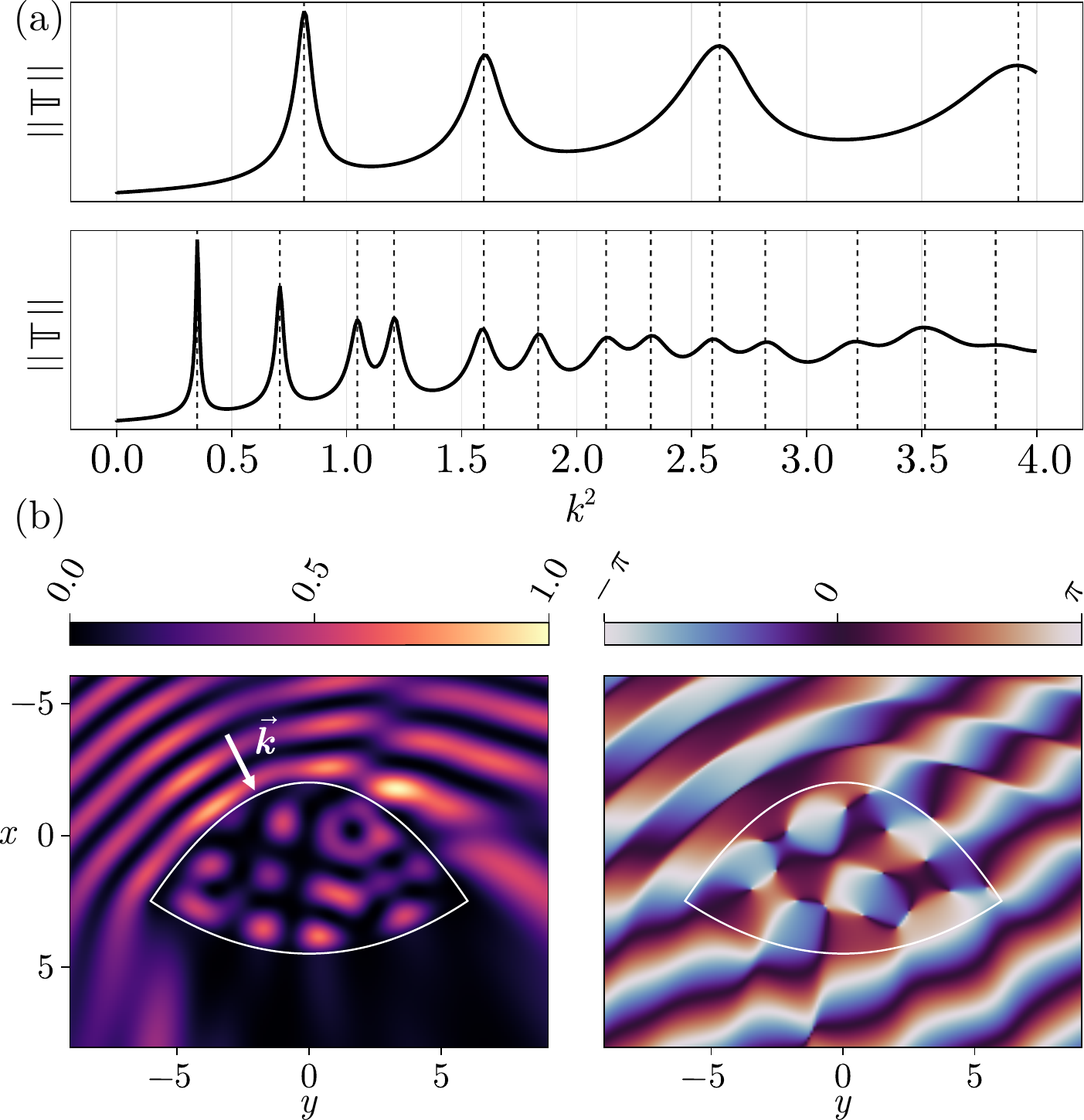}
    \caption{(a) Scan of the energy spectra of the billiards \parabolic{2}{2} (top) and \parabolic{3}{2} (bottom) for a \quote{leaky} boundary with $\gamma=2$. Compare this to \autoref{fig:resonances}. (b) Probability density and phase distribution of the scattered wavefield for an incident plane wave with $k=2.17$.}
    \label{fig:resonances:leaky}
\end{figure}

\vspace{3mm}
\textbf{Leaky boundary.}
The previous examples considered a $\delta$-like boundary with very high strength (infinite $\gamma$ approximation). To illustrate the case of a \textit{leaky} wall, Fig. \ref{fig:resonances:leaky}(a) shows the variation of $\norm{\mathbb{T}}$ in function of $k^2$ for the billiards analyzed in \autoref{fig:resonances} but now with lower potential given by $\gamma=2$. In this case, the matrix $\mathbb{T}$ has to be calculated with the expression without approximation \eqref{eqn:bwm:t:def}, rather than the simplified \eqref{eqn:t:limit}. Note that the resonance peaks are no longer as sharp and narrow as those shown in \autoref{fig:resonances}. In \autoref{fig:resonances:leaky}(b), we plot the wavefields scattered by the billiard \parabolic{3}{2} using $k=2.17$. Based on the phase distribution, we can conclude that the field inside the billiard is no longer a real function. Instead, it is a complex field that displays the wave's traveling behavior inside the billiard. We can also observe the presence of vortices in the phase. These phase singularities correspond to specific points inside the billiard where the probability density becomes zero, and the phase gets indeterminate.

%
\section{Conclusions}

We have solved the Lipmman-Schwinger equation for the scattering of plane waves by $\delta$-boundaries that have parabolic shapes and infinite or finite lengths. When the parabolic walls were open, we determined the solutions analytically in terms of several series of products of parabolic cylinder functions. We explored the solutions for both variable and constant strengths of the potential. To validate our analytic constructions, we compared them with the results obtained by the boundary wall method, corroborating an excellent agreement.

Our analysis focused on the resonance effect in confocal parabolic billiards with very high $\delta$ wall potential. We observed that stationary eigenstates of the billiard emerge when the wavenumber $k$ of the incident plane wave matches the eigenenergy $E_n=k_n^2$ of the confined particle in the parabolic billiard. Additionally, we corroborated that the angle of incidence of the incident plane wave plays an important role in tuning the states of the billiard. If the plane wave is incident along the billiard's symmetry axis, it is not possible to induce anti-symmetric eigenstates. To achieve this, we needed to break the symmetry of the problem by tilting the incident plane wave.

\begin{acknowledgments}
Alberto R. B. thanks Héctor Medel-Cobaxin and Raul I. Hernández-Aranda for helpful discussions about the curve discretization and the $\mathbb{T}$ matrix.
\end{acknowledgments}

\appendix

\section{Basic properties of the parabolic cylinder functions}\label{sec:appendix:weber}

In this appendix, we include some useful relations of the parabolic cylinder functions $D_{\nu}(z)$  that are useful for this work. An extensive treatment of the parabolic functions can be found in Refs. \cite{abramowitz1968,erdelyi,Gradshteyn,weberIdentities2016}.

The parabolic cylinder functions $D_{\nu}(z)$ are solutions of the parabolic cylinder differential equation
\begin{equation}\label{eqn:appendix:weber:eqn}
    \dv[2]{z}\weber{\nu}{z}+\qty(\nu+\frac{1}{2}-\frac{1}{4}z^2)\weber{\nu}{z}=0,
\end{equation}
and are directly related to the \textit{standard} parabolic cylinder functions $U(a,z)=\weber{-a-1/2}{z}$,
introduced by Whittaker \cite{whittaker1902,WhittakerandWatson} and standardized by \citet{miller1952}. Only two of the four solutions $D_{\nu}(z), D_{\nu}(-z), D_{-\nu-1}(\imag z)$, and $D_{-\nu-1}(-\imag z)$ are linearly independent.


Recursion relation
\begin{equation}
D_{\nu+1}(z)  -zD_{\nu}(z) +\nu D_{\nu-1}(z)=0.
\label{RCD}
\end{equation}

\textbf{Special values.}

For integer order $\nu=m\geq0$, the PCF reduces to Hermite-Gaussian functions
\begin{equation}\label{eqn:weber:hermite}
    \weber{m}{z}=2^{-m/2}e^{-z^2/4}\,H_m(z/\sqrt{2}),
\end{equation}
where $H_m(\cdot)$ is the Hermite polynomial of order $m$. The special case when $m=0$ corresponds to a Gaussian function
\begin{equation}
\label{ap:weber:0}
\weber{0}{z}=e^{-z^2/4}.
\end{equation}

If $\nu=-1$, the function $D_{\nu}$ reduces to
\begin{equation}
\label{ap:weber:1}
\weber{-1}{z}=e^{z^2/4}\sqrt{\frac{\pi}{2}}\erfc{\frac{z}{\sqrt{2}}},
\end{equation}
where $\erfc{\cdot}$ is the complementary error function \cite{abramowitz1968}.

The PCFs with negative integers $\nu=-m$, $m>1$, can be determined with the recursion relation Eq. (\ref{RCD}).
For example
\begin{equation}
D_{-2}(z)=e^{z^{2}/4}\sqrt{\frac{\pi}{2}}\left[  \sqrt{\frac{2}{\pi}}%
e^{-z^{2}/2}-z~\mathrm{erfc}\left(  \frac{z}{\sqrt{2}}\right)  \right],
\end{equation}
and so on.

\section{Boundary Wall Method} \label{appBWM}

The formal derivation of the BWM follows from \eqref{eqn:ls:delta} which, after writing the relationship between the incident wave-function $\phi(\vb{r}_a)$ and the scattered wave $\psi(\vb{r}')$ at two points in the boundary $\mathcal{C}$, i.e., $\vb{r}'_a$ and $\vb{r}'_b$, we can write~\cite{daluz1997quantum, zanetti2008resonant}
\begin{multline}
    \psi(\vb{r})=\phi(\vb{r})-\int_{\mathcal{C}}\int_{\mathcal{C}}  \dd{s_a}\dd{s_b}G_0(\vb{r},\vb{r}(s_b))\,\\
    \times\gamma(s_b)\,T_{\gamma}(s_b,s_a;k)\,\phi(\vb{r}_a),
\end{multline}
where $T_{\gamma}$ can be found from the recursive relationship
\begin{multline}\label{eqn:t_gamma}
    T_{\gamma}(s_b,s_a)=\delta(s_b-s_a)+\int_{\mathcal{C}}\dd{s_c}T_{\gamma}(s_b,s_c)\,\\\times\gamma(s)\,G_0(\vb{r}(s_c),\vb{r}(s_a)).
\end{multline}

The key of the method is that if $T_{\gamma}$ can be obtained, the wave function throughout space can be found by performing a definite integral over $\mathcal{C}$. Although $T_{\gamma}$ can sometimes be obtained from its definition, \eqref{eqn:t_gamma}, it is often challenging (some examples are given in \cite{zanetti2008eigenstates}). Instead, we approach the problem numerically, as explained below.

Following \citet{daluz1997quantum}, we define the $T$ operator as
\begin{equation}\label{eqn:bwm:t:def}
T= -\gamma T_{\gamma}=\gamma\qty[\widetilde{I}-\gamma \widetilde{G}_0]^{-1},
\end{equation}
whose action on the incident wave is given by
\begin{equation}\label{eqn:t_operator:action}
\gamma\widetilde\psi=T\widetilde\phi,
\end{equation}
where $\widetilde{I}$ is the identity operator, and we borrow their notation where the tilde implies that we are evaluating only at the boundary. The method can be extended to consider Neumann and mixed boundary conditions; however, these are not pertinent to our purposes. The numerical version treats a discrete form of the $T$-matrix seen above.

\medskip
\textbf{Matrix Treatment}

First, we partition the boundary $\mathcal{C}$ into $N$ equal segments $\qty{\mathcal{C}_j}$,
\begin{equation}\label{eqn:appendix:bwm:first}
    \psi(\vb{r})=\phi(\vb{r})+\sum_{j=1}^N\int_{\mathcal{C}_j}\dd{s}\,\gamma\,G_0(\vb{r},\vb{r}(s))\,\psi(\vb{r}(s)).
\end{equation}
Using a mean-value approximate, we write the wave function at the midpoint of each segment, $\vb{r}(s_j)$, which we call unambiguously $\vb{r}_j$,
\begin{equation}\label{eqn:bwm:meanval}
    \psi(\vb{r})\approx\phi(\vb{r})+\sum_{j=1}^N\psi(\vb{r}_j)\,\int_{\mathcal{C}_j}\dd{s}\,\gamma\,G_0(\vb{r},\vb{r}(s)).
\end{equation}
We then sets the observation point at another boundary segment $\vb{r}=\vb{r}_i$ to rewrite \eqref{eqn:appendix:bwm:first} into
\begin{equation}\label{eqn:bwm:approx}
    \psi(\vb{r}_i)\approx\phi(\vb{r}_i)+\sum_{j=1}^N\,\mathbb{M}_{ij}\,\gamma\,\psi(\vb{r}_j),
\end{equation}
where one defines $\mathbb{M}$ to be a matrix with entries
\begin{equation}
\mathbb{M}_{ij} \equiv \int_{\mathcal{C}_j}\dd{s}G_0(\vb{r}_i,\vb{r}(s)).
\end{equation}
Following this we define column vectors $\Psi$ and $\Phi$, each of length $N$, with respective elements $\qty{\psi(\vb{r}_j)}$ and $\qty{\phi(\vb{r}_j)}$; these are evaluated only at the boundary. We solve \eqref{eqn:bwm:approx}, obtaining
\begin{equation}\label{eqn:gammaPsi}
    \gamma\,\Psi_i=\gamma\sum_{j}^N\qty[\qty(\mathbb{1} - \gamma\,\mathbb{M})^{-1}]_{ij}\Phi_j=(\mathbb{T}\Phi)_i.
\end{equation}
with $\mathbb{T}$ being the discrete version of \eqref{eqn:bwm:t:def}. We can, therefore, insert this into \eqref{eqn:bwm:meanval} to obtain an expression for the wave function in the whole space,
\begin{equation}
    \psi(\vb{r})\approx\phi(\vb{r})+\sum_{j=1}^{N}G_0(\vb{r},\vb{r}_j)\deltaS{j}(\mathbb{T}\Phi)_j,
\end{equation}
where $\deltaS{j}$ is the length of $\mathcal{C}_j$ defined through a mean-value approximation of the integral in \eqref{eqn:bwm:meanval}. The diagonal elements $i=j$ are undefined due to the nature of the free-particle Green's function~\cite{zangwill2013modern}. Therefore, we have to perform the full integral at such elements~\cite{daluz1997quantum}.
\begin{equation}
\mathbb{M}_{ij}=\begin{dcases}
    \int_{\mathcal{C}_j}\dd{s'}\,G_0(\vb{r}_i,\vb{r}(s'))&i=j,\\
    G_0(\vb{r}_i,\vb{r}_j) \deltaS{j}&i\neq j.\\
\end{dcases}
\end{equation}

For the diagonal elements, we split the integral over the whole segment into two integrals over smaller segments, leaving the singularity $s_j$ at the endpoints, as seen in~\cite{maioli2018}. Using a mean-value approximation for each integral, we get
\begin{align}\label{eqn:bwm:double_integral}
\mathbb{M}_{ij}&=\int\limits_{s^-}^{s_j}\dd{s'}G_0(\vb{r}_i,\vb{r}(s'))+\int\limits_{s_j}^{s^+}\dd{s'}G_0(\vb{r}_i,\vb{r}(s')),\\
&\approx G_0(\vb{r}_i,\vb{r}'(s^-))\deltaS{j}^-+G_0(\vb{r}_i,\vb{r}'(s^+))\deltaS{j}^+,
\end{align}
with $\vb{r}(s^\pm)$ the vector position of the end-points of each segment, and $\deltaS{j}^\pm$ is the length of each smaller segment, $|s_j - s^\pm|$. Since $\vb{r}_i$ is defined as the midpoint of $\mathcal{C}_j$, then $\deltaS{j}^-=\Delta s_j^+=\deltaS{j}/2$, and therefore,

\begin{equation}
\mathbb{M}_{ii}\approx\alpha\hankel{k\deltaS{i}/2}\deltaS{i}.
\end{equation}
We find that using the \quote{fully approximated} $\mathbb{M}$ works well \cite{daluz1997quantum}.

Finally, we note the result from \citet{daluz1997quantum} where, for an \textit{impenetrable} barrier ($\gamma\rightarrow\infty$),
\begin{equation}\label{eqn:t:limit}
    \mathbb{T}=-(\mathbb{M}^{-1}).
\end{equation}

For a $\gamma(s)$ to vary along $\mathcal{C}$, a quick view of
the explicit form of the recursive relationship \eqref{eqn:t_gamma} seen above, shows that there is a $\gamma(s_j)$ factor multiplying each $G_0(\vb{r}_j, \vb{r}_{j-1})$~\cite{zanetti2008eigenstates},
\begin{multline}\label{eqn:recursive}
    T^{(j)}_{\gamma}(s_a,s_b)=\gamma^{j}\int\dd{s_1}\dots\dd{s_{j-1}}G_0(\vb{r}_b,\vb{r}_{j-1})\\
    \times G_0(\vb{r}_{j-1},\vb{r}_{j-2})\cdots G_0(\vb{r}_2,\vb{r}_1)~G_0(\vb{r}_1,\vb{r}_a),
\end{multline}
where $\vb{r}_j\equiv\vb{r}(s_j)$. Hence, we treat each $\gamma(s_j)$ as constant and incorporate its dependency on $s_j$ in \eqref{eqn:gammaPsi}. While we still have a Dirichlet boundary condition, $\psi(\vb{r})|_{\vb{r}\in\mathcal{C}}$ need not be zero.

\section{Discussion of the $\mathbb{T}$-matrix} \label{AppC}

\begin{figure}[t]
    \centering
    \includegraphics[width=\linewidth]{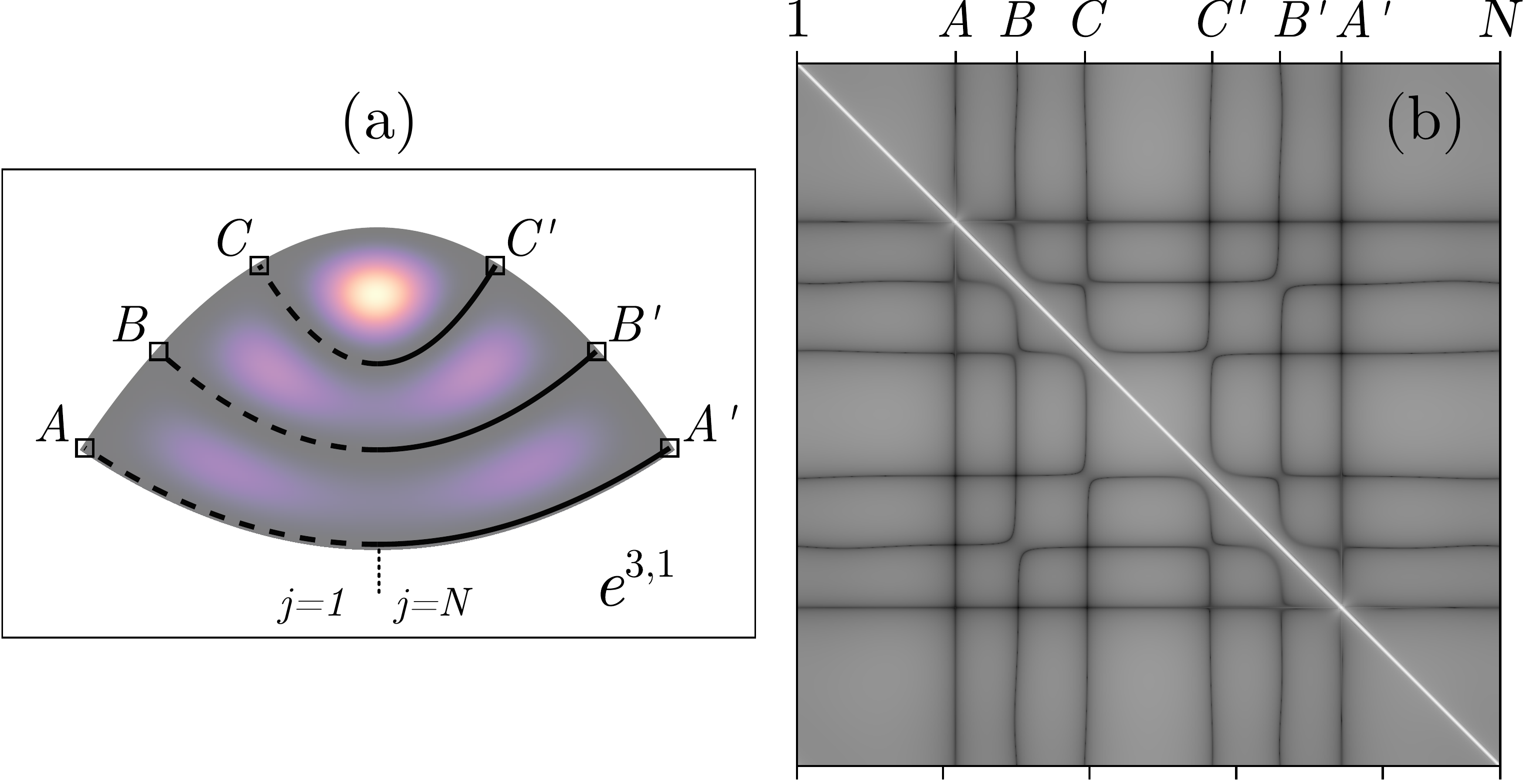}
    \caption{Correspondence between billiard's nodal lines from (a) $|\psi_n|^2$ and vertical locus (b) $ |\mathbb{T}|^2$ (log scale).}
    \label{fig:f14_correspondence}
\end{figure}

From \citet{zanetti2008eigenstates}, it is known that the \(\mathbb{T}\)-matrix acts as a propagator for $\phi(\vb{r}(s_a))$ to the segment $\vb{r}(s_b)$. Therefore, we have the following remark:

\begin{remark}\label{rem:prob}
    $|\mathbb{T}|^2$ considers the probability amplitude of a wave $\phi$ that impinges at $\vb{r}(s_a)$ and leaves at $\vb{r}(s_b)$.
\end{remark}

This is the reason why the elements near the main diagonal have higher intensity in \autoref{fig:tmatrix}, corresponding to specular reflection~\cite{zanetti2008eigenstates}. Vertical (or horizontal) lines of low probability amplitude, therefore, represent a \emph{homogeneous} behavior of \cref{rem:prob}. In other words, a vertical line at column $j$ represents an equally low probability of propagation from curve segment $\mathcal{C}_i$ to the rest of the billiard's segments (up to some band seen in \autoref{fig:tmatrix}).

A curved locus of (near) zeros simply means that, while the probability of propagation previously mentioned is low, it is not homogeneous with respect to the rest of the billiard. Furthermore, at an off-resonance $k$, one cannot talk about propagation by the $\mathbb{T}$-matrix, as the wave function vanishes inside the billiard.

There are two vertical (and, by symmetry, two horizontal) bands of low probability that appear in $|\mathbb{T}|^2$ throughout the spectrum. We attribute these regions to being close to the corners of the billiard; hence, these show up regardless of the geometry or $k$.

As one sweeps the spectrum $k$, whenever we reach an eigenenergy $k=k_n$, there are other (relatively) vertical lines that appear in the density plot for $|\mathbb{T}|^2$. We find a correspondence between the locus from $|\mathbb{T}|^2$ and the intersection between the billiard's boundary $\mathcal{C}$ and the nodal lines present in $|\psi_n|^2$, seen in \autoref{fig:f14_correspondence}. For more discussion on the periodic patterns of $\mathbb{T}$, we refer the reader to Ref. \cite{zanetti2008eigenstates}.

\end{document}